%% file: master.tex
\documentclass[a4paper,fleqn,10pt]{article}
\pdfoutput=1
\input{text/global}

\hypersetup{
  pdfauthor={Marek Schoenherr},
  pdftitle={An automated subtraction of NLO EW infrared divergences}
}
\preprint{CERN--TH--17--281\\MCnet--17--24}
\author{Marek Sch{\"o}nherr}
\title{An automated subtraction of NLO EW infrared \\[4mm] divergences}
\institute{Theoretical Physics Department, CERN, 1211 Geneva 23, Switzerland}
\begin{document}
\vspace*{10mm}
\maketitle
\vspace*{20mm}
\begin{abstract}
  In this paper a generalisation of the 
  Catani-Seymour dipole subtraction method to next-to-leading 
  order electroweak calculations is presented. 
  All singularities due to photon and gluon radiation off 
  both massless and massive partons in the presence of both 
  massless and massive spectators are accounted for. 
  Particular attention is paid to the simultaneous subtraction 
  of singularities of both QCD and electroweak origin which 
  are present in the next-to-leading order corrections to 
  processes with more than one perturbative order 
  contributing at Born level. 
  Similarly, embedding non-dipole-like photon splittings 
  in the dipole subtraction scheme discussed. 
  The implementation of the formulated subtraction scheme 
  in the framework of the \Sherpa Monte-Carlo event 
  generator, including the restriction of the dipole 
  phase space through the $\alpha$-parameters and 
  expanding its existing subtraction for NLO QCD 
  calculations, is detailed and
  numerous internal consistency checks validating 
  the obtained results are presented.
\end{abstract}
\newpage
\tableofcontents
\input{text/introduction}

\input{text/methods}

\input{text/implementation}

\input{text/results}
\input{text/conclusions}
\appendix
\input{text/appendix}

\bibliographystyle{amsunsrt_modpp}
\bibliography{journal}
  \end{document}

%% file: text/global.tex
\usepackage{amsmath}
\usepackage{amssymb}
\usepackage{array}
\usepackage{calc}
\usepackage{longtable}
\usepackage{multirow}
\usepackage{pstricks}
\usepackage{graphicx}
\usepackage{xspace}
\usepackage{units}

\numberwithin{equation}{section}
\usepackage{mciteplus}
\usepackage[pdfborder={0 0 0}]{hyperref}
\usepackage[format=hang,labelfont=bf,hypcap=true]{caption}
\usepackage{subcaption}
\usepackage{sectsty}
\usepackage{enumitem}
\allsectionsfont{\sffamily}
\subsubsectionfont{\mdseries\itshape\large}
\makeatletter
\DeclareRobustCommand*{\bfseries}{%
  \not@math@alphabet\bfseries\mathbf
  \fontseries\bfdefault\selectfont
  \boldmath
}
\makeatother
\let\spreprint\empty
\newcommand{\preprint}[1]{\def\spreprint{\protect#1}}
\let\sinstitute\empty
\newcommand{\institute}[1]{\def\sinstitute{\protect#1}}
\makeatletter
\renewcommand{\maketitle}{\begingroup
  \null\thispagestyle{empty}%
    \ifx\spreprint\empty
      \vskip 5ex
    \else
      \flushright\large\spreprint\vskip 10ex
    \fi
    \vskip 5ex
    \flushleft
      {\sffamily\bfseries\huge\@title}\vskip 6ex
      \@author\vskip 2ex
      \ifx\sinstitute\empty
      \else
        {\small\sinstitute}
      \fi
    \vskip 5ex
  \endgroup
}
\makeatother
\renewenvironment{abstract}{\begin{center}
  {\large\sffamily\bfseries Abstract: }
  \begin{minipage}[t]{0.75\textwidth}
}{\end{minipage}\end{center}\vskip 10ex}

\numberwithin{equation}{section}
\allowdisplaybreaks[2]

\newcommand{\LHAPDF}{L\protect\scalebox{0.8}{HAPDF}\xspace}
\newcommand{\MRSTqed}{M\protect\scalebox{0.8}{RST}qed\xspace}

\newcommand{\CTqed}{CT14qed\xspace}
\newcommand{\LUXqed}{L\protect\scalebox{0.8}{UX}qed\xspace}
\newcommand{\NNPDFtwoqed}{N\protect\scalebox{0.8}{NPDF}2.3qed\xspace}
\newcommand{\NNPDFthreeqed}{N\protect\scalebox{0.8}{NPDF}3.0qed\xspace}
\newcommand{\NNPDFthreeoneqed}{N\protect\scalebox{0.8}{NPDF}3.1luxqed\xspace}
\newcommand{\MSbar}{\ensuremath{\overline{\text{MS}}}\xspace}

\newcommand{\Rivet}{R\protect\scalebox{0.8}{IVET}\xspace}

\newcommand{\GoSam}{G\protect\scalebox{0.8}{O}S\protect\scalebox{0.8}{AM}\xspace}

\newcommand{\Recola}{R\protect\scalebox{0.8}{ECOLA}\xspace}
\newcommand{\OpenLoops}{O\protect\scalebox{0.8}{PEN}L\protect\scalebox{0.8}{OOPS}\xspace}
\newcommand{\FastJet}{F\protect\scalebox{0.8}{AST}J\protect\scalebox{0.8}{ET}\xspace}
\newcommand{\Sherpa}{S\protect\scalebox{0.8}{HERPA}\xspace}
\newcommand{\Comix}{C\protect\scalebox{0.8}{OMIX}\xspace}

\newcommand{\Amegic}{A\protect\scalebox{0.8}{MEGIC}\xspace}

\long\def\symbolfootnote[#1]#2{\begingroup%
\def\thefootnote{\fnsymbol{footnote}}\footnote[#1]{#2}\endgroup}

\newcommand{\im}{\imath}
\newcommand{\jm}{\jmath}
\newcommand{\ijt}{{\widetilde{\im\hspace*{-1pt}\jm}}}

\newcommand{\kt}{{\tilde{k}}}
\newcommand{\ajt}{{\widetilde{a\;\!\!\jm}}}
\newcommand{\at}{{\tilde{a}}}
\newcommand{\bt}{{\tilde{b}}}
\newcommand{\qt}{{\tilde{q}}}
\newcommand{\cykt}{c_\kt^\gamma}

\newcommand{\done}{{\rm d}}
\newcommand{\order}{\mathcal{O}}
\newcommand{\alphaS}{\alpha_s}
\newcommand{\alphaQED}{\alpha}
\newcommand{\alphadip}{\alpha_\text{dip}}
\newcommand{\alphaFF}{\alpha_\text{\tiny FF}}
\newcommand{\alphaFI}{\alpha_\text{\tiny FI}}
\newcommand{\alphaIF}{\alpha_\text{\tiny IF}}
\newcommand{\alphaII}{\alpha_\text{\tiny II}}
\newcommand{\mc}[1]{\mathcal{#1}}
\newcommand{\mr}[1]{\mathrm{#1}}

\newcommand{\mf}[1]{\mathbf{#1}}

\newcommand{\bea}{\begin{eqnarray}}
\newcommand{\eea}{\end{eqnarray}}
\newcommand{\bi}{\begin{itemize}}
\newcommand{\ei}{\end{itemize}}

\newcommand{\mHl}{\vphantom{\int\limits_A^B}}
\newcommand{\Kallen}[3]{\lambda\left(#1,#2,#3\right)}
\newcommand{\DiLog}[1]{\text{Li}_2\left(#1\right)}
\newcommand{\Iop}{\boldsymbol{I}}
\newcommand{\Kop}{\boldsymbol{K}}
\newcommand{\Pop}{\boldsymbol{P}}
\newcommand{\Qop}[1]{\boldsymbol{Q}_{#1}^2}
\newcommand{\nspec}{n_\text{spec}}
\newcommand{\Kbar}{\overline{K}}
\newcommand{\KFS}{K_\text{FS}}
\newcommand{\Kcal}{\mc{K}}
\newcommand{\Kt}[1]{K_{#1}^t}
\newcommand{\Ktilde}{\tilde{K}}
\newcommand{\Preg}[1]{P_\text{reg}^{#1}}
\newcommand{\Phatp}[1]{\hat{P}^{\prime #1}}
\newcommand{\V}{\mathcal{V}}
\newcommand{\VS}{\V^\text{(S)}}
\newcommand{\VNS}{\V^\text{(NS)}}
\newcommand{\nlf}{{\ensuremath{n_\text{lf}}}}
\newcommand{\nf}{{\ensuremath{n_\text{f}}}}

\newcommand{\TR}{{\ensuremath{T_R}}}
\newcommand{\NC}{{\ensuremath{N_C}}}

\newcommand{\muR}{\mu_\text{R}}
\newcommand{\muF}{\mu_\text{F}}

\newcommand{\shortequal}{\!\!=\!\!}
\newcommand{\sigmaIRD}{\sigma_\text{IRD}}

\newlist{myitemize}{itemize}{3}
\setlist[myitemize]{leftmargin=14em}


%% file: text/introduction.tex
\section{Introduction}

As Run-I of the LHC has been successfully completed, 
culminating in the celebrated experimental confirmation 
of the existence of the Higgs boson, Run-II proceeds 
its data-taking at the unprecedented centre-of-mass 
energy of 13\,TeV. 
As the much anticipated discovery of signals of 
beyond-the-Standard-Model physics is still lacking, 
precision tests scrutinising the Standard Model are 
of prime importance, now and in the foreseeable future. 
At the same time, new physics searches are looking 
for increasingly small signals demanding more precise 
estimates of the Standard Model backgrounds. 
This expansion of sensitivity of both precision 
measurements and new physics searches in the multi-TeV 
region demand an immense improvement of theoretical 
predictions. 

This precision can be achieved by the inclusion of 
next-to and next-to-next-to-leading order (NLO and NNLO) 
corrections in the strong coupling and next-to-leading 
order electroweak (EW) corrections. 
Here it should be noted that both NNLO QCD and NLO EW 
corrections are expected to be of a similar magnitude 
for inclusive observables as numerically 
$\alphaS^2\approx\alpha$. 
On selected differential distributions, however, 
electroweak corrections can grow much larger. 
They are dominated by photon emissions in the distributions 
of final state leptons, for example. 
In invariant mass spectra of lepton pairs below a 
resonance, for example, $O(1)$ corrections can be present, 
in which case a proper resummation should be included 
\cite{Yennie:1961ad}. 
Similarly, looking at the (multi-)TeV regime, 
the NLO EW corrections quickly grow considerably, 
reducing cross sections by a few tens of percent, 
due to the emergence of large electroweak Sudakov 
corrections arising as the scattering energies 
$Q^2\gg m_W^2$ 
\cite{
  Sudakov:1954sw,
  Beenakker:1993tt,Beenakker:1993rm,Beenakker:1993yr,
  Denner:1996ug,
  Fadin:1999bq,Kuhn:1999nn,Ciafaloni:2000df,Hori:2000tm,
  Denner:2000jv,Denner:2001gw,Melles:2001dh,Beenakker:2001kf,
  Feucht:2004rp,Jantzen:2005xi,Baur:2006sn}.
In this regime they are larger than even the 
NLO QCD corrections in many cases and their omission 
becomes the dominant uncertainty in experimental 
studies and searches. 

To this end, benefiting from the well-established 
techniques developed for the automation of NLO QCD 
corrections many NLO EW corrections have been 
calculated recently 
\cite{
  Denner:2014ina,Kallweit:2014xda,Denner:2014bna,
  Kallweit:2015fta,Frixione:2015zaa,Chiesa:2015mya,
  Denner:2015fca,Kallweit:2015dum,
  Biedermann:2016yvs,Biedermann:2016guo,
  Denner:2016jyo,Frederix:2016ost,Denner:2016wet,
  Kallweit:2017khh,Lindert:2017olm,Granata:2017iod,
  Chiesa:2017gqx,Biedermann:2017bss,Biedermann:2017oae,
  Greiner:2017mft,Frederix:2017wme}. 
To fully automate these computations at NLO EW 
accuracy in a Monte-Carlo framework all infrared 
divergences need to be regulated, where various 
incarnations of subtraction methods have proven 
to be the methods of choice for practical 
implementations \cite{Frixione:1995ms,Catani:1996vz,
Catani:2002hc,Kosower:1997zr,Nagy:2003qn}.
Similar subtractions have also been published 
for NLO EW calculations using a dipole picture 
\cite{Yennie:1961ad,Dittmaier:1999mb,Dittmaier:2008md,Gehrmann:2010ry}. 
Contrary to the QCD case, only the implementation of 
\cite{Gehrmann:2010ry}, restricted to photon 
emissions of fermions, is publicly available though.

Besides the generalisation to all divergent splittings 
at $\order(\alpha)$, including photon splittings and 
photon emissions off massive scalars and vector bosons, 
this publication addresses the issue 
of automatically detecting simultaneously occurring 
QCD and QED singularities and subtracting them consistently. 
These occur as soon as the Born process is defined at multiple 
orders $\order(\alphaS^n\alpha^{N-n})$. 
In this case the NLO EW correction to the $\order(\alphaS^n\alpha^{N-n})$ 
process, being of $\order(\alphaS^n\alpha^{N-n+1})$, 
is at the same time the NLO QCD correction to 
the $\order(\alphaS^{n-1}\alpha^{N-n+1})$ process and will 
in general exhibit the corresponding infrared singularities. 
Further, matters of the organisation of the contributing 
partonic processes and their mapping to reduce 
the computational complexity along with the provision 
of infrared safe phase space cuts are discussed. 
The algorithm is implemented in the \Amegic \cite{Krauss:2001iv} 
matrix element generator which is part of the \Sherpa 
\cite{Gleisberg:2008ta} Monte-Carlo event generator 
framework. 
It bases on the automated subtraction of massless NLO QCD 
divergences therein \cite{Gleisberg:2007md,Archibald:2011nca}.
The implementation presented in this publication has, 
in various preliminary forms, already been used to 
calculate electroweak corrections to a multitude of 
important signal and background processes 
\cite{
  Kallweit:2014xda,Kallweit:2015fta,Kallweit:2015dum,Badger:2016bpw,
  Alioli:2016fum,Biedermann:2017yoi,
  Kallweit:2017khh,Lindert:2017olm,Chiesa:2017gqx,Greiner:2017mft}, 
highlighting its versatility.

The present paper is structured as follows: 
First, in Section \ref{sec:method} the Catani-Seymour 
dipole subtraction method is reviewed and the general 
modifications to its differential and integrated 
subtraction terms are discussed. 
Section \ref{sec:imp} then details its automation 
in \Sherpa's matrix element generator \Amegic, 
highlighting the necessary changes and improvements 
with respect to \cite{Gleisberg:2007md,Archibald:2011nca}. 
This section also discusses various options 
implemented for the incorporation of photon splittings, 
general infrared safe fiducial phase space definitions 
and flavour scheme conversions. 
Essential cross checks validating the presented 
implementation are then presented in Section \ref{sec:results} 
before concluding in Section \ref{sec:conclusions}.
Explicit formulae for all differential and integrated 
dipoles are given in Appendix \ref{app:diffsplit}--\ref{app:Ialpha}.

%% file: text/methods.tex
\section{Catani-Seymour subtraction at NLO EW}
\label{sec:method}

In order to be applicable to NLO EW calculations the well-known 
Catani-Seymour dipole subtraction \cite{Catani:1996vz,Catani:2002hc} 
needs to be recast in a suitable form. 
To highlight the changes from the original formulation 
for NLO QCD calculations the complete structure of the formalism 
is reviewed. 
This subtraction formalism starts from the 
the expectation value of any infrared safe observable $O$ described at NLO 
accuracy through
\begin{equation}
  \label{eq:nlo}
  \begin{split}
    \langle O\rangle^\text{NLO}
    \,=\;&
      \int\done\Phi_m^{(4)}\;\mr{B}(\Phi_m^{(4)})\,O(\Phi_m^{(4)})\\
    &{}
      +\left[
	\int\done\Phi_m^{(d)}\;
	\left[
	  \mr{V}(\Phi_m^{(d)})
	  +\mr{C}(\Phi_m^{(d)})
	\right] O(\Phi_m^{(d)})
        +\int\done\Phi_{m+1}^{(d)}\;
	  \mr{R}(\Phi_{m+1}^{(d)})\;O(\Phi_{m+1}^{(d)})
       \right]_{\epsilon=0}\;.
  \end{split}
\end{equation}
Therein, the Born term $\mr{B}$ consist of the squared matrix element 
$\left|\mc{M}_m\right|^2={}_m\langle s_1,\ldots,s_m|s_1,\ldots,s_m\rangle{}_m$ 
with helicity states $s_n$ and further includes all parton densities, 
parton fluxes, symmetry and averaging factors. 
The virtual and real corrections, $\mr{V}$ and $\mr{R}$, as well as the 
collinear counter term, $\mr{C}$, are defined analogously. 
When regulating their respective divergences through dimensional 
regularisation, they have to be evaluated consistently in $d=4-2\epsilon$ 
dimensions for all singularities to cancel. 
Only after their summation can the limit $\epsilon\to 0$ be taken. 
$\done\Phi_n^{(4)}$ and $\done\Phi_n^{(d)}$ are the 
four and $d$ dimensional phase space element. 
As $\mr{V}$ and $\mr{C}$ on the one hand side and $\mr{R}$ on the 
other are defined on phase spaces of different parton multiplicity, 
eq.\ \eqref{eq:nlo} cannot be used for numerical evaluation straight 
forwardly and it is rewritten as 
\begin{equation}
  \label{eq:cs-nlo}
  \begin{split}
    \langle O\rangle^\text{NLO}
    \,=\;&
      \int\done\Phi_m^{(4)}\;\mr{B}(\Phi_m^{(4)})\,O(\Phi_m^{(4)})\\
    &{}
      +\int\done\Phi_m^{(4)}\;
	\left[
	  \mr{V}(\Phi_m^{(d)})
	  +\mr{C}(\Phi_m^{(d)})
	  +\int\done\Phi_1^{(d)}\,\mr{D}(\Phi_m^{(d)}\cdot\Phi_1^{(d)})
	\right]_{\epsilon=0} O(\Phi_m^{(4)})\\
    &{}
      +\int\done\Phi_{m+1}^{(4)}
	\left[
	  \mr{R}(\Phi_{m+1}^{(4)})\;O(\Phi_{m+1}^{(4)})
	  -\mr{D}(\Phi_m^{(4)}\cdot\Phi_1^{(4)})\;O(\Phi_m^{(4)})
	\right] \,,
  \end{split}
\end{equation}
introducing the subtraction term $\mr{D}$. 
For its construction it is a mandatory requirement that 
$\mr{D}\to\mr{R}$ in all singular limits, rendering 
the integral on the third line finite in four dimensions. 
The divergences of the virtual correction and the collinear 
counterterm on the other side are cancelled separately by 
the integral of the subtraction term over the one-particle phase 
space, rendering the integrals of the second line 
finite in four dimensions as well. 
Sec.\ \ref{sec:method:diffsub} now describes the construction 
of the differential subtraction term, $\mr{D}$, used to 
subtract all divergences from the real emission correction, 
while Sec.\ \ref{sec:method:intsub} presents the integrated 
subtraction terms, $\mr{I}_\mr{D}=\int\done\Phi_1\;\mr{D}$, 
that subtracts all divergences from the virtual corrections 
and the collinear counterterm in its explicit Laurent expansion 
after being analytically integrated over the factorised 
one-particle phase space.

All infrared divergences that occur at NLO EW are of QED origin. 
No subtractions of potentially large, but finite, corrections 
involving the emissions of real and virtual massive electroweak 
gauge bosons will be considered. 
The practical implementation described in Section \ref{sec:imp} 
follows the general lines of \cite{Gleisberg:2007md,Archibald:2011nca}.

\subsection{Differential subtraction terms}
\label{sec:method:diffsub}

\begin{figure}[t!]
  \begin{center}
    \includegraphics{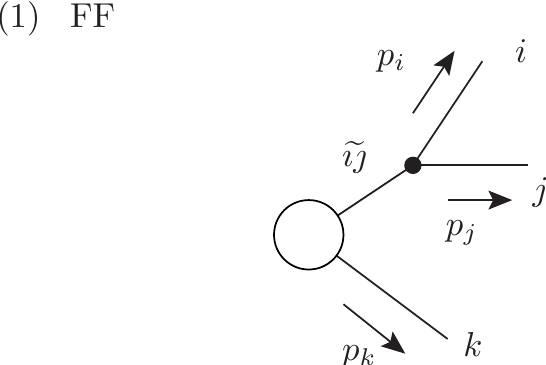}\hspace*{20mm}
    \includegraphics{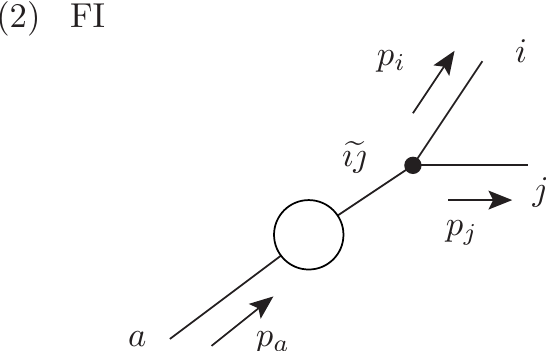}\\
    \includegraphics{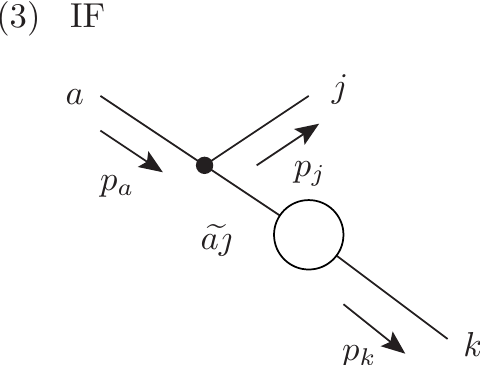}\hspace*{26.5mm}
    \includegraphics{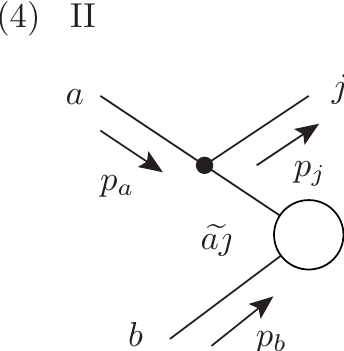}\hspace*{19.5mm}
  \end{center}
  \caption{
    Classification of the four dipole types in Catani-Seymour-type 
    dipole subtraction. 
    \label{fig:diptypes}
  }
\end{figure}

To describe all singular limits of a given real emission matrix 
element $\mc{R}$, related to the real emission term of eq.\ \eqref{eq:nlo} 
in a similar fashion as the Born term, it is decomposed into 
a sum over dipoles $\mc{D}$ \cite{Catani:1996vz,Catani:2002hc} as
\begin{equation}
  \label{eq:dipole_factorisation}
  \begin{split}
    \left|\mc{M}_{n+1}\right|^2=\mc{R}
    \,\to\;&\mc{D}=
          \sum\limits_{i,j}\sum\limits_{k\neq i,j}\mc{D}_{ij,k}
	  +\sum\limits_{i,j}\sum\limits_{a}\mc{D}_{ij}^a
	  +\sum\limits_{a,j}\sum\limits_{k\neq j}\mc{D}_{j,k}^{a}
	  +\sum\limits_{a,j}\sum\limits_{b\neq a}\mc{D}_j^{a,b}\;.
  \end{split}
\end{equation}
Therein, $i$ is the emitter in the final state, $j$ is the emittee, $k$ 
is the spectator in the final state, $a$ and $b$ are the initial state 
partons. 
Each dipole thus encodes the singularity structure caused by the emission 
of $j$ in the presence of the charge of the spectator. 
While the divergence associated with the soft emission of $j$ off the 
dipole $\ijt-\kt$ is partially fractioned into a piece associated with a
splitting $\ijt\to i+j$ in the presence of spectator $k$ and a piece 
where $i$ and $k$ swap their roles, the divergence associated with 
the collinear emission of $j$ off $\ijt$ is recovered through charge 
conservation once all dipoles having $\ijt$ as emitter are summed over. 

All four dipole types are depicted in Fig.\ \ref{fig:diptypes}.
The individual dipoles take the form 
\cite{Catani:1996vz,Catani:2002hc,Nagy:1998bb,Nagy:2003tz}
\begin{equation}
  \begin{split}
    \mc{D}_{ij,k}
    \,=\;& -\frac{1}{(p_i+p_j)^2-m_\ijt^2}\;\Qop{\ijt\kt}\;
	   {}_m\langle\ldots,\ijt,\ldots,\kt,\ldots|\mf{V}_{ij,k}|
	              \ldots,\ijt,\ldots,\kt,\ldots\rangle{}_m\;
	    \Theta(\alphaFF-y_{ij,k})\\
    \mc{D}_{ij}^a
    \,=\;& -\frac{1}{(p_i+p_j)^2-m_\ijt^2}\,\frac{1}{x_{ij,a}}\;\Qop{\ijt\at}\;
	   {}_m\langle\ldots,\ijt,\ldots,\at,\ldots|\mf{V}_{ij}^a|
	              \ldots,\ijt,\ldots,\at,\ldots\rangle{}_m\;
	    \Theta(\alphaFI-1+x_{ij,a})\\
    \mc{D}_{j,k}^{a}
    \,=\;& -\frac{1}{2p_ap_j}\,\frac{1}{x_{aj,k}}\;\Qop{\ajt\kt}\;
	   {}_m\langle\ldots,\ajt,\ldots,\kt,\ldots|\mf{V}_{j,k}^a|
	              \ldots,\ajt,\ldots,\kt,\ldots\rangle{}_m\;
	    \Theta(\alphaIF-u_j)\\
    \mc{D}_j^{a,b}
    \,=\;& -\frac{1}{2p_ap_j}\,\frac{1}{x_{aj,b}}\;\Qop{\ajt\bt}\;
	   {}_m\langle\ldots,\ajt,\ldots,\bt,\ldots|\mf{V}_j^{a,b}|
	              \ldots,\ajt,\ldots,\bt,\ldots\rangle{}_m\;
	    \Theta(\alphaII-v_j)\;.
  \end{split}
\end{equation}
Therein, the charge-correlator is defined as \cite{Yennie:1961ad,Dittmaier:1999mb,Dittmaier:2008md}
\begin{equation}\label{eq:chargecorrelator}
  \begin{split}
    \Qop{\ijt\kt}
    \,=\,
    \left\{\begin{array}{cl}
	  \frac{Q_\ijt Q_\kt \theta_\ijt \theta_\kt}{Q_{\ijt}^2} & \ijt\neq\gamma\\
	  \kappa_{\ijt\kt} & \ijt=\gamma
    \end{array}\right.
    \qquad\qquad\text{and}\qquad\qquad
    \sum_{\kt\neq \ijt}\kappa_{\ijt\kt}\,=\;-1\quad\forall\ijt=\gamma.
  \end{split}
\end{equation}
The $Q_\ijt$ and $Q_\kt$ are the charges of the emitter and the 
spectator and their $\theta_{\ijt/\kt}$ are $1$ ($-1$), if they are in the 
final (initial) state. 
Of course, $Q_\kt=Q_k$ and $\theta_\kt=\theta_k$.
In the case of photon splittings no soft divergence is 
present. 
Thus, these splittings have no dipole character. 
To include them in the dipole-formalism nonetheless and to distribute 
the recoil in splittings away from the collinear limit, spectator 
partons need to be assigned. 
As their only role is to absorb transverse momentum of the splitting 
process, any other particle may be considered as spectator. 
Each thus assigned recoil partner may be assigned a weight 
$\kappa_{\ijt\kt}$, only their sum is constraint by 
eq.\ \eqref{eq:chargecorrelator} in order to add up to the correct 
collinear limit. 
Various options to assign recoil partners are implemented, 
they are detailed in Sec.\ \ref{sec:photonrecoil}.

Since the QED charges are real numbers, the charge-correlator 
simply multiplies the matrix element and only leaves the 
spin-correlators $\mf{V}_{ij,k}$, $\mf{V}_{ij}^a$, $\mf{V}_{j,k}^a$ 
and $\mf{V}_j^{a,b}$ as insertions in the spin-correlated underlying 
Born matrix elements. 
The spin-correlators directly correspond to their QCD counterparts 
and are detailed in Appendix \ref{app:diffsplit}. 
It also defines the initial state momentum rescaling parameters 
$x_{ij,a}$, $x_{aj,k}$ and $x_{aj,b}$, as well as the splitting variables 
$y_{ij,k}$, $u_j$ and $v_j$. 
The $\{\alphadip\}=\{\alphaFF,\alphaFI,\alphaIF,\alphaII\}$ parameters 
serve to restrict the phase space where the individual dipole 
terms are non-zero and therefore need to be evaluated 
\cite{Nagy:1998bb,Nagy:2003tz}. 
They are constructed such that for every $\alpha_{\ijt\kt}>0$ the 
singularity is fully subtracted. 
The introduction of a parameter $\kappa$ in dipoles where a final 
state photon splits into a massive fermion pair in the presence 
of a final state spectator similarly allows a redistribution of 
finite terms, cf.\ Appendix \ref{app:diffsplit}.

\subsection{Integrated subtraction terms}
\label{sec:method:intsub}

By the above construction, the integral of the subtraction 
terms $\mr{D}$ over the one-particle phase space possesses all 
the necessary poles to render the second line in 
eq.\ \eqref{eq:cs-nlo} finite as $\epsilon\to 0$.
This section now summarises the formulae and findings of 
\cite{Catani:1996vz,Catani:2002hc}, 
translated to the QED case, and discusses their important features.
The integrated subtraction terms, together with the collinear counterterm, 
are commonly reorganised into $\Iop$, $\Kop$ and $\Pop$ operators through 
the following identification
\begin{equation}\label{eq:def_IKP}
  \begin{split}
    \lefteqn{\hspace*{-5mm}
      \sum_{a,b}\int\done\eta_a\done\eta_b
      \int\done\Phi_m^{(4)}\;
	\left[
	  \mr{V}_{ab}(\Phi_m^{(d)})
	  +\mr{C}_{ab}(\Phi_m^{(d)})
	  +\int\done\Phi_1^{(d)}\,\mr{D}_{ab}(\Phi_m^{(d)}\cdot\Phi_1^{(d)})
	\right]_{\epsilon=0} O(\Phi_m^{(4)})
      }\\
    \,=\;&
      \sum_{a,b}\int\done\eta_a\done\eta_b
      \int\done\Phi_m^{(4)}\;
      \Bigg\{
	\left[
	  \mr{V}_{ab}(\Phi_m^{(d)})
	  +\mr{B}_{ab}(\Phi_m^{(d)})\cdot\Iop(\epsilon,\mu^2;\kappa,\{\alphadip\})
	\right]_{\epsilon=0}\\
    &\hspace*{38mm}{}
      +\sum_{a'}\int\done x_a\;
        \mr{B}_{a'b}(\Phi_m^{(4)})\cdot
	\left[
	  \Kop_{aa'}(x_a;\{\alphadip\})
	  +\Pop_{aa'}(x_a;\mu_F^2)
	  \vphantom{\Phi_m^{(4)}}
	\right]\\
    &\hspace*{38mm}{}
      +\sum_{b'}\int\done x_b\;
        \mr{B}_{ab'}(\Phi_m^{(4)})\cdot
	\left[
	  \Kop_{bb'}(x_b;\{\alphadip\})
	  +\Pop_{bb'}(x_b;\mu_F^2)
	  \vphantom{\Phi_m^{(4)}}
	\right]
      \Bigg\}\; O(\Phi_m^{(4)})\;.\hspace*{-10mm}
  \end{split}
\end{equation}
The $\Iop$ operator contains the necessary infrared poles to cancel 
all divergences of the virtual correction such that, after summing 
both, the limit $\epsilon\to 0$ can be taken and the integral can 
be evaluated in four dimensions. The $\Kop$ and $\Pop$ operators 
are infrared finite by construction and, thus, evaluation in four 
dimension is unproblematic as well. Please note, 
that due to the fact that, contrary to the colour correlator in QCD, 
the charge correlator in QED is a simple real number and thus the 
convolution of the Born matrix element with the respective insertion 
operators in QCD becomes a trivial product of the Born matrix 
element and the operators in QED. The spin-correlation that was still 
present in the real subtraction terms has been integrated out in 
full analogy to the QCD case. In the following, the structure 
of all three operators in the QED case is discussed.

\paragraph*{The $\protect\Iop$ operator.}

The $\Iop$ operator contains all flavour-diagonal endpoint contributions and 
cancels all divergences present in the one-loop matrix elements. It 
takes the general form
\begin{equation}\label{eq:I}
  \begin{split}
    \Iop(\epsilon,\mu^2;\kappa,\{\alphadip\})
    \,=\;&-\frac{\alphaQED}{2\pi}\,\frac{(4\pi)^\epsilon}{\Gamma(1-\epsilon)}\;
	  \sum_i\sum_{k\neq i}\; \Iop_{ik}(\epsilon,\mu^2;\kappa,\{\alphadip\})
  \end{split}
\end{equation}
and contains single and, in case of massless emitters, double poles 
in $\epsilon$. 
Due to the presence of such poles a dependence on the regularisation 
scale $\mu^2$ enters. 
It is commonly identified with the renormalisation scale $\muR^2$. 
The $\Iop$ operator is further dependent on the choice of the 
$\{\alphadip\}$ and $\kappa$ parameters that effect non-singular 
terms only. 
It is further decomposed into dipoles \cite{Catani:2002hc}
\begin{equation}\label{eq:Iik}
  \begin{split}
    \Iop_{ik}(\epsilon,\mu^2;\kappa,\{\alphadip\})
    =\;&
	\Qop{ik}
	\left[
	  \V_{ik}(\epsilon,\mu^2;\kappa)
	  +\Gamma_i(\epsilon,\mu^2)
	  +\gamma_i\left(1+\ln\frac{\mu^2}{s_{ik}}\right)
	  +K_i
	  +A_{ik}^I(\{\alphadip\})
	  +\order(\epsilon)
	\right]\,,\hspace*{-10mm}
  \end{split}
\end{equation}
wherein $i$ labels the emitter and $k$ the spectator. The charge insertion 
operator, which is a trivial real number in the QED case, is defined in 
eq.\ \eqref{eq:chargecorrelator}.
The full crossing invariance of the QCD $\Iop$ operator may be broken 
in the QED case 
in the presence of photon splittings as their recoil partner assignment 
is arbitrary and may involve information on initial or final state 
particles. 
Different possible choices are discussed in Sec.\ \ref{sec:photonrecoil}, 
some of which may break this crossing invariance.

The divergences of the $\Iop$ operator are encoded in the functions 
$\V_{ik}$ and $\Gamma_i$. While the former contains all 
soft-(quasi-)collinear divergences the latter contains the pure 
(quasi-)collinear ones. They do not only differentiate whether 
the emitter is a photon or not, but also between different spins 
of the emitter. 
Their precise form as well as the the flavour constants $\gamma_i$ 
and $K_i$ are given in Appendix \ref{app:intsplit}. 
$A_{ik}^I$ encodes the dependence on the phase space restriction 
of the individual dipoles $\{\alphadip\}$. 
Finite terms originating in dipoles involving initial state legs, 
however, can be pushed into the $\Kop$ operator. 
Thus, $A_{ik}^I$ by convention only depends on $\alphaFF$. 
Its precise form is given in Appendix \ref{app:Ialpha}.

\paragraph*{The $\protect\Kop$ and $\protect\Pop$ operators.}

The $\Kop$ and $\Pop$ operators collect all pieces of the 
integrated dipole terms that are not collected in the $\Iop$ 
operator and combines them with the collinear counterterms 
$\mr{C}$ to give a finite result as $\epsilon\to 0$. 
By construction they contain only remainders of splittings 
where either the emitter or the spectator is in the initial 
state. 
Thus, they are comprised of terms arising due to the change 
of the flavour or the partonic momentum fraction $x$ of an 
initial state due to a splitting.

The $\Kop$-operator is given by \cite{Catani:2002hc}
\begin{equation}
  \label{eq:Kop}
  \begin{split}
    \Kop_{aa'}(x;\{\alphadip\})
    \,=\;&
      \frac{\alpha}{2\pi}
      \left\{
	\Kbar_{aa'}(x)
	-\KFS(x)
	-\sum_i\Qop{ia'}\Kcal_{i,aa'}(x)
	-\sum_k\Qop{a'k}\Kt{aa',k}(x)
      \right.\\
    &\hphantom{\frac{\alpha}{2\pi}\;\;}\left.{}
	-\Qop{a'b}\Ktilde_{aa'}(x)
	+A_{aa'}^K(\{\alphadip\})
	\vphantom{\sum_k}
      \right\}\;.
  \end{split}
\end{equation}
It depends on the partonic $x$, and the flavour change 
from the Born process initial state flavour $a$ to $a'$
of the convolution eq.\ \eqref{eq:def_IKP}. 
Therein, the $\Kbar$ collect universal terms present 
in all splitting involving an initial state as either 
emitter or spectator. 
Then, while $\Kcal$ contains solely remaining terms from final 
state splittings in the presence of initial state spectators, 
the $\Kt{}$ are their counterparts for initial state splittings 
in the presence of a final state spectator, 
$i$ and $k$ running over all final state partons in each case.
$\Ktilde$ contains solely related correlations between both 
initial states, arising from dipoles where both the emitter 
and the spectator are in the initial state. 
The $A^K$ terms collect all finite terms arising when any 
of $\alphaFI$, $\alphaIF$ or $\alphaII$ is different from unity, 
thus restricting the phase space of the respective dipoles. 
Again, the $\Qop{ik}$ are the charge correlators of 
eq.\ \eqref{eq:chargecorrelator}.
Finally, $\KFS$ contains the factorisation scheme dependence. 
Currently, both only the \MSbar schemes is supported, 
setting these terms identically zero. 

The $\Pop$-operator now collects the remaining initial state collinear 
singularity from all dipoles involving initial states either 
as emitters or as spectators and cancels them against the collinear 
counterterm. Through this counterterm a dependence on the 
factorisation scale enters. 
The $\Pop$ operator is given by \cite{Catani:1996vz,Catani:2002hc}
\begin{equation}
  \label{eq:Pop}
  \begin{split}
    \Pop_{aa'}(x,\mu_F^2)
    \,=\;&
      \frac{\alpha}{2\pi}\;P^{aa'}(x)
      \left[
	\sum\limits_k\Qop{a'k}\,\log\frac{\muF^2}{xs_{ak}}
	+\Qop{a'b}\,\log\frac{\muF^2}{xs_{ab}}
      \right]\;.
  \end{split}
\end{equation}
Only initial state splittings are present, either in the 
presence of a spectator in the final state, which is encoded in 
the sum of $k$, or with the opposite initial state $b$ acting as 
the spectator. 
It otherwise only depends on the Alterelli-Parisi splitting 
function detailed in Appendix \ref{app:intsplit}.

%% file: text/implementation.tex
\section{Implementation}
\label{sec:imp}

The implementation of the QED generalisation of the Catani-Seymour 
dipole subtraction scheme in \Sherpa's matrix element generator 
\Amegic proceeds along the lines of \cite{Gleisberg:2007md}. 
As in general real and virtual corrections of $O(\alphaS^n\alpha^m)$ 
contain divergences of both QCD and QED origin, both cases are 
included in this section.
In the following, the general structure of the implementation is 
reviewed. 

\subsection{Identification of dipoles}
\label{sec:builddip}

The starting point to construct the involved subtraction terms 
in the Catani-Seymour subtraction formalism is a given flavour 
configuration in the Born or the real emission phase space and 
the perturbative order $\order(\alphaS^n\alpha^m)$ in accordance 
with the respective virtual or real correction to be computed. 
For all parts, on-the-fly variations of both the factorisation 
scale $\muF$ and the renormalisation scale $\muR$ are available 
through an extension of the algorithm detailed in \cite{Bothmann:2016nao}.

\paragraph{Differential subtraction terms.}
A given real emission configuration $\{ab\}\to\{1,..,m+1\}$ at order 
$O(\alphaS^n\alpha^m)$ can in general exhibit both QCD and QED 
divergences simultaneously. 
The following therefore describes the identification of both types of dipoles.
Thus, all triplets $\{i,j,k\}$ that can be built from the 
external particles of the process are tested for the presence of 
an infrared divergence, QCD or QED, by checking for the existence 
of a dipole subtraction term.
In these triplets $i$ and $k$ may be in the initial or final state 
while $j$ may be in the final state only. 
Likewise, $i\neq j$, $i\neq k$, $j\neq k$ and triplets that only differ 
in a permutation of $i$ and $j$ are considered identical.
Then, the following steps are executed.
\begin{enumerate}[label=\arabic*)]
  \item Based on the quantum numbers and flavours of the triplet it is 
        decided whether a QCD, a QED or both splitting function can exist. 
        A QCD splitting function can exist only if $i$, $j$ and $k$ are 
        colour charged, while a QED splitting function can exist only if 
        the charge-correlator $\Qop{\ijt\kt}$ does not vanish. 
        The dipole type is determined based on whether $i$ and $k$ 
        are in the initial or final state. 
        A given triplet $\{i,j,k\}$ may exhibit both QCD and QED 
        divergences, and thus may form both a QCD and a QED dipole.
  \item The flavours $\ijt$ and $\kt$ are determined for each possible 
        splitting function.
  \item For each possible splitting function $\{\ijt,\kt\}\to\{i,j,k\}$ 
        the underlying Born configuration and its order, 
        $O(\alphaS^{n-1}\alpha^m)$ in case of a QCD splitting function 
        and $O(\alphaS^n\alpha^{m-1})$ in case of a QED splitting 
        function, are determined. 
        If, including the insertion of the appropriate colour-, charge- 
        and spin-correlations, such a process at this order exists, 
        a dipole subtraction term is built.
\end{enumerate}
If the above steps do not lead to a valid dipole subtraction term, 
no divergence can be present.
The real emission, a conventional tree-level process, is grouped 
together with all its subtraction terms into one computational 
unit and their respective cross sections and observable values, 
$O(\Phi_{m+1})$ and all $O(\Phi_{m_i})$, are treated as correlated.

\paragraph{Integrated subtraction terms.}
Similar to the above discussed real emission corrections, 
the virtual correction configurations $\{ab\}\to\{1,..,m\}$ 
at order $O(\alphaS^n\alpha^m)$ in general exhibits poles due 
to both QCD and QED origins. 
To subtract them, both QCD and QED integrated subtraction 
need to be included. 
In fact, they naturally arise as counterparts to differential 
subtraction terms constructed for the corresponding real 
emission correction, as guaranteed by Bloch-Nordsieck 
\cite{Bloch:1937pw} or Kinoshita-Lee-Nauenberg 
\cite{Kinoshita:1962ur,Lee:1964is} theorem. 
Consequently, QCD and QED $\Iop$, $\Kop$ and $\Pop$ operators 
are constructed. 
While their QCD version are described in detail in 
\cite{Gleisberg:2007md,Archibald:2011nca}, their QED version of eq.\ \eqref{eq:def_IKP} 
are discussed below.

The $\Iop$ operator, on the one hand side, has the same initial 
state flavours and momentum fractions as the virtual correction. 
The $\Kop$ and $\Pop$ operators on the other hand, 
resulting from the combination of the integrated subtraction 
terms and the collinear counterterms, involve a summation over 
possible initial state flavours and 
comprise the following general structure in their dependence on 
the additional $x$ integration variable 
\begin{equation}
  \begin{split}
    \left[g(x)\right]_++\delta(1-x)h(x)+k(x)\;.
  \end{split}
\end{equation}
Therein, both $h(x)$ and $k(x)$ are regular functions in $x$, 
while the plus distribution of $g(x)$ is defined as
\begin{equation}
  \begin{split}
    \int\limits_0^1\done x\;f(x)\left[g(x)\right]_+
    \,=\;&
      \int\limits_0^1\done x\;\left[f(x)-f(1)\right] g(x)\;.
  \end{split}
\end{equation}
Hence, the potentially computationally intensive matrix 
elements in $\mr{B}$ have to be calculated twice for 
every phase space point in addition to the flavour 
summation. 
To remedy this, the original expression of eq.\ \eqref{eq:def_IKP} 
is reformulated. 
Dropping the dependence on the $\{\alphadip\}$ parameters and 
explicitly stating the dependence of the underlying Born term 
on the initial state momentum fractions and parton densities, 
$\mr{B}_{ab}=f_af_b\,\mc{B}_{ab}$, the $\Kop$ and $\Pop$ 
operators it can be recast to
\begin{equation}
  \begin{aligned}[b]
    \lefteqn{\hspace*{-5mm}
      \sum_{a,b}\int\done\eta_a\done\eta_b
      \int\done\Phi_m^{(4)}\;
        \sum_{a'}\int\limits_0^1\done x_a\;
        \mr{B}_{a'b}(x_a\eta_a,\eta_b;\Phi_m^{(4)})\cdot
	\left[
	  \Kop_{aa'}(x_a;\{\alphadip\})
	  +\Pop_{aa'}(x_a;\mu_F^2)
	  \vphantom{\Phi_m^{(4)}}
	\right]
      \;O(\Phi_m^{(4)})
    }\\
    =\;&
      \sum_{a,b}\int\done\eta_a\done\eta_b
      \int\done\Phi_m^{(4)}\;
        \sum_{a'}\int\limits_0^1\done x_a
	\left\{
	  g^{aa'}(x_a)
	  \left[
	    \mr{B}_{a'b}(x_a\eta_a,\eta_b;\Phi_m^{(4)})
	    -\mr{B}_{a'b}(\eta_a,\eta_b;\Phi_m^{(4)})
	  \right]\vphantom{\int}
	\right.\\
    &\hspace*{53mm}\left.{}
	  +k^{aa'}(x_a)\,
	    \mr{B}_{a'b}(x_a\eta_a,\eta_b;\Phi_m^{(4)})
	  +h^{aa'}(1)\,
	    \mr{B}_{a'b}(\eta_a,\eta_b;\Phi_m^{(4)})\vphantom{\int}
	\right\}
      \;O(\Phi_m^{(4)})\hspace*{-15mm}\\
    =\;&
      \sum_{a,b}\int\done\eta_a\done\eta_b
      \int\done\Phi_m^{(4)}\;
      f_b(\eta_b)\,
      \mc{B}_{ab}(\Phi_m^{(4)})\\
    &{}
      \times\sum_{a'}
      \left\{\;
	\int\limits_{\eta_a}^1\done x_a
	\left[
	  \tfrac{1}{x_a}\,f_{a'}\!\!\left(\tfrac{\eta_a}{x_a}\right)
	  \left(g^{aa'}(x_a)+k^{aa'}(x_a)\right)
	  -f_{a'}(\eta_a)\,g^{aa'}(x_a)
	\right]
      \right.\\
    &\hspace*{12mm}
      \left.{}\vphantom{\frac{f_{a'}\left(\frac{\eta_a}{x_a}\right)}{x_af_a(\eta_a)}}
	+f_{a'}(\eta_a)
	 \left(
	   h^{aa'}-G^{aa'}(\eta_a)
	 \right)
      \right\}
      \;O(\Phi_m^{(4)})\;,
  \end{aligned}
\end{equation}
with $f_a(\eta)$ being the parton density of flavour $a$ and 
momentum fraction $\eta$ in the proton, otherwise absorbed in 
the Born term $\mr{B}$. 
All PDFs are evaluated at the same scale $\muF$. 
$G^{ab}(\eta)=\int_0^\eta\done x\;g^{ab}(x)$ are the analytically 
computed divergence free parts of the integral of the functions 
under the plus distribution. 
Its divergence at $x=1$ is cancelled numerically on the second last line. 
Effectively, this reformulation results in a redefinition of the 
PDF for incoming parton $a$ of $\mr{B}_{ab}$. 
As the integrand of the remaining integral over $x_a$ is a simple 
one-dimensional function without a pronounced structure, its numerical 
evaluation is very stable and can be accomplished by a 
single point for each summand per phase space point $\done\Phi_m^{(4)}$. 
The $\Kop$ and $\Pop$ operators for the second incoming parton $b$ 
are recast similarly.

The thus transformed form of the $\Kop$ and $\Pop$ operators require 
only a single evaluation of the potentially costly matrix elements 
in $\mr{B}$ while retaining the number of computations of PDFs 
needed, speeding up the computation considerably for involved 
processes. 
This allows to generate and evaluate the underlying Born matrix 
element for both the $\Iop$ operators and the $\Kop$ and $\Pop$ 
operators at the same time. 
In fact, due to these three operators being simple multiplicative 
scalars, the common underlying Born matrix element is identical 
to the standard Born matrix element, allowing for their simultaneous 
calculation at no extra cost.
The operators themselves are then built from dipoles constructed 
from all doublets $\{i,k\}$, $i\neq k$, of external partons 
available in the partonic process for which the charge-correlator 
$\Qop{ik}$ does not vanish.

\subsection{External photons}
\label{sec:photonrecoil}

External photons can play different roles in a calculation:
they can either be resolved or unresolved. 
According to this distinction they should be treated 
differently at NLO EW \cite{Harland-Lang:2016lhw,Kallweit:2017khh}. 
Initial state photons are always unresolved at 
a hadron collider. 
They thus should be treated in a short-distance scheme, 
allowing them to split into massless fermions, necessitating 
a proper subtraction of infrared divergences. 
Final state photons on the other hand can play both roles. 
If they are considered resolved, they should be treated in an 
on-shell scheme and no explicit photon splitting is allowed. 
Concerning the dipole subtraction discussed in this paper 
they are thus neutral particles and do not form part of a 
dipole, except as possible recoil partner of another unresolved 
photon. 
A final state unresolved photon, on the other hand, again must 
be treated in a short-distance scheme, necessitating their splittings 
to be subtracted.

As discussed in Section \ref{sec:method:diffsub}, the dipole 
picture is not necessary to capture the divergences of photons 
splitting in pairs of massless fermions, leptons and quarks, due 
to the absence of soft divergences that necessitate the 
correlation with the emissions off other partons of the event. 
Nonetheless, it offers a possibility to assign one or more 
recoil partners to absorb the transverse momentum of the 
splitting, thus fitting these purely collinear splittings 
into the dipole picture. 
The choice of spectator is essentially arbitrary, and all 
other partons of the event offer being viable spectators. 
Only the following condition has to hold
\begin{equation}\label{eq:kappa-photon}
  \begin{split}
    \sum_{\kt\neq \ijt}\kappa_{\ijt\kt}
    \,=\;-1
  \end{split}
\end{equation}
for every splitting photon $\ijt$, cf.\ \ref{eq:chargecorrelator}.
Therein, the $\kappa_{\ijt\kt}$ are arbitrary and possibly dynamic 
weights assigned to every dipole with spectator $\kt$. 
In practise, for initial state as well as final state 
splittings five choices $\cykt$ have been implemented:
\begin{enumerate}
  \setcounter{enumi}{-1}
  \item only allow initial state partons as spectators, 
  \item only allow final state partons as spectators,
  \item only allow QED charged particles as spectators, 
  \item only allow QED neutral particles as spectators, 
  \item allow all particles as spectators.
\end{enumerate}
The $\kappa_{\ijt\kt}$ are set to the phase space point independent 
value $-\nspec^{-1}$, with $\nspec$ the number of assigned spectators, 
thus trivially fulfilling eq.\ \eqref{eq:kappa-photon}.

\subsection{Process mappings}
\label{sec:method:map}

In general, physical cross sections include multiple different 
partonic channels. 
However, many of these partonic channels share identical squared 
matrix elements, potentially differing by constant factors. 
They, thus, do not need to be recomputed for every flavour 
channel but can be reused. 
As a typical example, consider the production of a lepton pair 
in association with two jets. 
The process $g\,d\to e^+e^-\,g\,d$ shares a common squared 
matrix element with $g\,s\to e^+e^-\,g\,s$ and $g\,b\to e^+e^-\,g\,b$ 
on Born level at $\order(\alphaS^2\alpha^2)$. 
Hence, the squared matrix elements of the latter two partonic 
processes are mapped on the first, reusing its computed value. 
In this way, of the 95 partonic channels of this process, only 
30 have to be computed. 
Further mappings of individual graphs and subgraphs are implemented 
but are not discussed in the following, see 
\cite{Krauss:2001iv,Gleisberg:2008ta}.

In \Sherpa's matrix element generators various forms of process mappings 
are implemented to reduce the computational complexity and memory footprint, 
both for the virtual and real emission corrections, 
cf.\ \cite{Krauss:2001iv,Gleisberg:2008ta,Gleisberg:2007md,Archibald:2011nca}. 
However, while for NLO QCD corrections to the leading order Born process 
it is true that if the Born process is mappable onto another existing 
process, then also both the virtual correction and the 
insertion-operator-augmented colour-correlated underlying Born process of 
the integrated subtraction term are mappable to the same process. 
This is no longer true when considering the NLO EW or
NLO QCD corrections to subleading Born processes.

In the present implementation of a full NLO QCD and NLO EW subtraction 
in the matrix element generator \Amegic the following process mapping 
strategy is followed for the real subtracted contributions:
\begin{itemize}
  \item The real emission process and its associated dipole subtraction 
        terms are grouped in one computational unit. 
        A given partonic channel of the real emission process can be mapped 
        onto another already existing one if both processes consist 
        of the same diagrams and all involved (internal and external) 
        particles have the same masses and widths and the same underlying 
        interaction (coupling factors may differ by a constant).
        Is this the case the whole computational unit can be mapped 
        and the result of the mapped-to process can simply 
        be reused.
  \item Individual dipoles can be mapped if the emitter, emittee and 
        spectator indices are identical, and the underlying Born 
        process can be mapped according to the above rules. 
        In this case, the result of the mapped-to process can simply 
        be reused.
  \item Underlying Born matrix elements can be mapped if the Born-level 
        emitter $\ijt$ carries the same indices and the underlying Born 
        process itself can be mapped. 
        The spin correlation insertion operator, needed if 
        parton $\ijt$ is either a gluon or a photon and described in 
        \cite{Gleisberg:2007md}, is encoded in the calculational 
        routines and necessitates the above restriction. 
        Can the underlying Born process be mapped, only the calculational 
        routines can be shared, reducing the memory footprint, but due to 
        the potentially 
        differing underlying Born momenta its result has to be recomputed.
\end{itemize}

The strategy for the virtual subtracted contributions reads 
as follows:
\begin{itemize}
  \item The virtual correction process, interfaced from an external 
        virtual correction provider, 
        as well as its associated integrated dipole subtraction terms 
        and the collinear counterterm, taken together and reformulated 
        according to Section \ref{sec:builddip}, are grouped 
        into one computational unit. 
        If the underlying Born processes of the integrated subtraction 
        contribution (in all orders required) can be mapped 
        according to the above rules and the virtual correction 
        provider confirms that the virtual correction can be mapped 
        onto the same process that \Sherpa's tree-level matrix element 
        generator maps the underlying Born 
        processes onto, the whole computational unit is mapped. 
        In this case, the result of the mapped-to process can simply 
        be reused.
        The $\Kop$ and $\Pop$ operators, whose internal PDF factors depend on the 
        initial state flavour, still have to be recomputed, but their matrix 
        element coefficients are cached.
  \item If the virtual correction provider cannot confirm the mapping 
        performed for the underlying Born process, only the underlying 
        Born process is mapped and virtual correction is recomputed.
        Again, the $\Kop$ and $\Pop$ have to be recomputed, but their matrix 
        element coefficients are cached. 
        Here, efficiency is lost if the virtual correction provider 
        uses less efficient process mappings.
\end{itemize}

\subsection{Fiducial phase space definition}
\label{sec:cuts}

Phase space restrictions are an essential part of the implementation 
of a framework for automated NLO calculcations. 
These cuts, however, need to be applied in an infrared safe way. 
At NLO, they must not discriminate between a massless parton before and 
after its collinear splitting or before and after a soft gluon or 
photon emission. 
Thus, if QCD singularities are present, massless QCD charged particles 
must be clustered into jets before any further cuts are applied. 
Similarly, in case of the presence of QED divergences, massless charged 
particles must either be dressed with the surrounding photons 
or be included in the jet algorithm. 
Subtleties arise in the presence of both QCD and QED singularities 
simultaneously. 
Here, usually, only a fully democratic jet finding can consistently 
treat all singularities, although specialised solutions exist for 
simplified situations.
Massive QCD and QED charged particles may be treated as bare as their 
mass shields the collinear singularity, but can also be included 
into jet finding and dressing algorithms.
An intermediate scheme which includes only the logarithms of the 
parton mass needed to regulate the collinear divergences 
\cite{Dittmaier:2008md}, but are otherwise treated massless in the 
calculation, is not implemented.

To this end, the implementation in \Sherpa is equipped with a range 
of algorithms to define infrared safe quantities on which further 
restrictions can be applied. Multiple such selectors can be nested.
\begin{myitemize}
  \item[\texttt{DressedParticleSelector}]
    This selector takes a choice of dressing algorithm (cone or sequential 
    recombination) and (flavour-dependent) dressing parameters 
    (cone radius or radial parameter and exponent). 
    All charged particles of the process are then dressed with 
    all photons using the specified algorithm with the given 
    (flavour-dependent) parameters. 
    Their four momenta are added such that four momentum is conserved. 
    The dressed charged particles may no longer be on-shell and the 
    photons used to dress the charged particles are removed from the 
    list of particles. 
    The resulting list of particles and their momenta are then 
    passed to all subselectors.
  \item[\texttt{Jet\_Selector}]
    This selector uses \FastJet \cite{Cacciari:2011ma} to build jets 
    from a given list of input particles. 
    It takes a list of flavours that are considered as jet finding 
    input particles, the jet finding algorithm and its parameters 
    including phase space boundaries in $p_\perp$, $\eta$ or $y$, 
    as well as a minimal and maximal number of jets to be found 
    as arguments.
    Additionally, clustered jets can be tagged or anti-tagged based 
    on their flavour content, including relative and absolute 
    constituent momentum requirements (e.g.\ $b$ tagging a jet 
    if one of its constituents is a $b$-quark or anti-$\gamma$-tagging 
    a jet if its photon constitents carry in excess of $z_\text{thr}$ 
    fraction of the total jet momentum). 
    The clustered jets as well as all particles not used as jet 
    finding input are passed on to all subselectors.
  \item[\texttt{Isolation\_Selector}]
    This selector uses the smooth cone isolation of \cite{Frixione:1998jh} 
    to isolate particles of a given flavour against particles 
    of another flavour. 
    It takes both the isolation flavour and rejection flavour list as 
    well as the algorithm parameters as input. 
    It further can be specified how many isolated particles of the 
    given flavour should minimally and maximally be found in a 
    given phase space volume (bounded by $p_\perp$ and $\eta$ or 
    $y$ ranges), e.g.\ exactly two isolated photons with 
    $p_\perp>30\,\text{GeV}$ and $|\eta|<2.35$. 
    The list of found isolated flavours and all other particles 
    except for those that should be isolated, but are not, 
    are passed to all subselectors.
\end{myitemize}
Once an infrared safe definition of particles is found, the standard 
single- or multi-particle selectors such as \texttt{PT} (implementing 
transverse momentum requirements on a given flavour), \texttt{PTmis} 
(missing transverse momentum build from all neutrinos), or \texttt{DR} 
(angular distance between the two given particles) can be used.
Further, \Sherpa is equipped with a user hook system, providing 
the possibility for users to implement arbitrary routines for 
phase space cuts and dynamically load them at run time, without 
the need to modify their \Sherpa installation.

Identified particles should in principle be defined using fragmentation 
functions, denoted $D_i^j(z)$ for finding parton $i$ in parton $j$ at 
momentum fraction $z$. 
As, however, all $D_i^i(z)$ have a $\delta(1-z)$ as leading term 
in on-shell renormalisation schemes (and only differ by ratios 
of couplings in other schemes) simplified schemes exist that are 
applicable to many practical situations. 
Hence, no fragmentation function is implemented yet.

\subsection{Flavour scheme conversion}
\label{sec:method:flavscheme}

All publicly available PDF sets containing QED effects in their 
evolution are fitted with five (\MRSTqed \cite{Martin:2004dh}, 
\CTqed \cite{Schmidt:2015zda}, \LUXqed \cite{Manohar:2016nzj,Manohar:2017eqh}, 
HKR16 \cite{Harland-Lang:2016apc},
\NNPDFthreeqed \cite{Ball:2014uwa}, \NNPDFthreeoneqed \cite{Bertone:2017bme}) 
or six (\NNPDFtwoqed \cite{Ball:2013hta}) light flavours. 
Thus, for next-to-leading order calculations with four or less light flavours the 
following scheme-conversion terms need to be added for consistency 
\cite{Cacciari:1998it}
\begin{equation}
  \begin{split}
    \lefteqn{\hspace*{-5mm}\langle O\rangle_\text{NLO QCD}^{(\nlf)}(\muR^2,\muF^2)}\\[2mm]
    \;=&\;
	\langle O\rangle_\text{NLO QCD}^{(\nf)}(\muR^2,\muF^2)\\[2mm]
	  &{}
	+ \int\!\done\Phi_m
	  \sum\limits_{i=\nlf}^\nf\sum_{\{ab\}}
	  \frac{\alphaS}{3\pi}\,
	  \TR
	  \left[
	    p\,\log\frac{m_i^2}{\muR^2}\,\Theta\left(\muR^2-m_i^2\right)
	    -\Delta_{ab}^{gg}\log\frac{m_i^2}{\muF^2}\,
	     \Theta\left(\muF^2-m_i^2\right)
	  \right]
	  \mr{B}_{ab}^{(\nf)}(\Phi_m;\muR^2,\muF^2)\;O(\Phi_m)
	  \hspace*{-20mm}
  \end{split}
\end{equation}
and
\begin{equation}
  \begin{split}
    \lefteqn{\hspace*{-5mm}\langle O\rangle_\text{NLO EW}^{(\nlf)}(\muR^2,\muF^2)}\\[2mm]
    \;=&\;
	\langle O\rangle_\text{NLO EW}^{(\nf)}(\muR^2,\muF^2)\\[2mm]
	  &{}
	- \int\done\Phi_m
	  \sum\limits_{i=\nlf}^\nf\sum_{\{ab\}}
	  \frac{\alpha}{3\pi}
	  \NC\,Q_i^2\;
	  \Delta_{ab}^{\gamma\gamma}\log\frac{m_i^2}{\muF^2}\,
	  \Theta\left(\muF^2-m_i^2\right)\;
	  \mr{B}_{ab}^{(\nf)}(\Phi_m;\muR^2,\muF^2)\;O(\Phi_m)
	  \;.
  \end{split}
\end{equation}
Therein, the $\langle O\rangle_\text{NLO QCD/EW}^{(n)}$ is the 
expectation value of an arbitrary observable $O$ computed at NLO QCD 
or NLO EW, respectively, with $n$-flavour scheme parton densities 
and the corresponding strong 
coupling summed over all initial state contributions. 
$\mr{B}_{ab}^{(n)}$ is the Born term, as defined in Section \ref{sec:method},  
in the $ab$ channel. The sums run over all $(\nf-\nlf)$ flavours of 
mass $m_i$ that are part of the \nf-flavour PDF parametrisation but not 
the \nlf-flavour scheme of the calculation and all combinations of 
partonic channels $ab$ occurring in the \nlf-flavour scheme, respectively. 
$p$ is the power of the strong coupling in the Born process and 
$\Delta_{ab}^{gg}=\delta_{ga}+\delta_{gb}$, i.e.\ it takes the 
values 2 in the $gg$ channel, 1 in all $qg$ and $\bar{q}g$ channels, and 
zero otherwise. $\Delta_{ab}^{\gamma\gamma}$ is defined analogously. 
Each logarithm of course only contributes if the scale is larger than the 
respective mass. As the electroweak coupling is not taken running in the 
common renormalisation schemes it is independent of the number of massless 
flavours in the calculation. For \MSbar{}-like renormalisation schemes a 
similar term proportional to its power in the Born process is to be added.

%% file: text/results.tex
\section{Checks of the implementation}
\label{sec:results}

\begin{table}[tb]
  \begin{center}
    \begin{tabular}{rclrcl}
      $G_\mu$ & \shortequal & $1.1663787\cdot 10^{-5}~\text{GeV}^2$ & \qquad\qquad & &\\
      $m_W$ & \shortequal & $80.385~\text{GeV}$  & $\Gamma_W$ & \shortequal & $2.0897~\text{GeV}$ \\
      $m_Z$ & \shortequal & $91.1876~\text{GeV}$ & $\Gamma_Z$ & \shortequal & $2.4955~\text{GeV}$ \\
      $m_t$ & \shortequal & $173.2~\text{GeV}$   & $\Gamma_t$ & \shortequal & $1.339~\text{GeV}$ \\
    \end{tabular}
  \end{center}
  \caption{
    Numerical values of all input parameters. The gauge boson masses are 
    taken from \cite{Agashe:2014kda}, while their widths are obtained from 
    state-of the art calculations. The Higgs mass and width are taken from 
    \cite{Heinemeyer:2013tqa}. The top quark mass is taken from 
    \cite{Agashe:2014kda} while its width has been calculated at NLO QCD.
    In calculations where a massive particle is present as an external 
    state, its width is set to zero.
    \label{tab:inputs}
  }
\end{table}

The implementation of the formalism described in the previous sections 
in \Sherpa needs to be validated. 
To this end, both its independence of its internal free parameters, 
$\{\alphadip\}=\{\alphaFF,\alphaFI,\alphaIF,\alphaII\}$, $\kappa$ 
and the choice of spectator in photon splittings, 
and its agreement with independent implementations for fixed 
values of these parameters need to be tested. 
While the latter were carried out in \cite{Kallweit:2014xda,
  Kallweit:2015fta,Kallweit:2015dum,Kallweit:2017khh,Lindert:2017olm} 
against the private implementations in \textsc{Munich} \cite{KallweitMunich}, 
in \cite{Badger:2016bpw} against \textsc{MadGraph5} \cite{Alwall:2014hca}, 
and in \cite{Badger:2016bpw,Biedermann:2017yoi} against 
\textsc{Recola} \cite{Actis:2012qn} and found full agreement, 
the former represent a powerful check of internal consistency. 
Further, as the $\{\alphadip\}$ parameters regulate the phase space 
coverage of the differential subtraction terms, they can be used to 
lower the average number of contributing dipoles for a given real 
emission contribution and, thus, reduce the computational costs of 
the real-subtracted contribution.

There are now two independent aspects of the calculation that need 
to be checked: 
\begin{itemize}
  \item[a)] The $\Iop$ operator of Section \ref{sec:method:intsub} 
            containing the explicit Laurent expansion in $\epsilon$ 
            as $\epsilon\to 0$ needs to reproduce the correct 
            $\epsilon^{-2}$ and $\epsilon^{-1}$ coefficients in 
            order to cancel all corresponding poles of the 
            virtual matrix elements, leading to a finite integrand 
            in eq.\ \eqref{eq:cs-nlo} in $d=4$. 
            These checks were performed for all possible dipoles 
            in \cite{Kallweit:2014xda,Kallweit:2015fta,Kallweit:2015dum,
              Kallweit:2017khh,Lindert:2017olm,Chiesa:2017gqx,
              Greiner:2017mft, Badger:2016bpw,Alioli:2016fum,
              Biedermann:2017yoi}
            and are not repeated here.
  \item[b)] The expectation value of any infrared observable is independent 
            of the choice of the technical $\{\alphadip\}$ and 
            $\kappa$ parameters as well as the choice $\cykt$ of 
            spectator for photon splittings. 
            It thus needs to be 
            verified that the sum of the contributions of the $\epsilon^0$ 
            coefficient of the integrated subtraction terms and 
            the differential subtraction terms is independent 
            of these parameters and choices. 
            In order to arrive at a finite result in $d=4$, 
            cf.\ eq.\ \eqref{eq:cs-nlo}, the real emission 
            correction and the collinear counterterms are added. 
            The corresponding quantity is defined as 
	    \begin{equation}
	      \label{eq:def-IRD}
	      \begin{split}
		\langle O\rangle_\text{IRD}
		\,=\;&
		  \int\done\Phi_m^{(4)}\;
		    \left[
		      \mr{I}_\mr{D}(\Phi_m^{(4)};\{\alphadip\},\kappa,\cykt)
		      +\mr{C}(\Phi_m^{(4)})
		    \right]_{\epsilon^0\,\text{coeff.}} O(\Phi_m^{(4)})\\
		&{}
		  +\int\done\Phi_{m+1}^{(4)}
		    \left[
		      \mr{R}(\Phi_{m+1}^{(4)})\;O(\Phi_{m+1}^{(4)})
		      -\mr{D}(\Phi_m^{(4)}\cdot\Phi_1^{(4)};\{\alphadip\},\kappa,\cykt)\;O(\Phi_m^{(4)})
		    \right] \,,
	      \end{split}
	    \end{equation}
	    and will be evaluated for processes containing all available 
	    dipole configurations in the following.
\end{itemize}
In the following, the inclusive or fiducial cross section contribution 
$\sigmaIRD$ is evaluated for a range of different processes, 
testing all possible dipole configurations. 
Throughout the input parameters of Table \ref{tab:inputs} are used 
in the $G_\mu$ scheme, although several other EW input parameter 
schemes are available in general. 
Similarly, all unstable particles are treated in the complex-mass scheme 
\cite{Denner:2005fg}.
Further, the CT14nlo \cite{Dulat:2015mca} and CT14qed \cite{Schmidt:2015zda} 
\footnote{
  To be precise the \texttt{CT14nlo} and \texttt{CT14qed\_inc\_proton} 
  PDF sets interfaced through \LHAPDF \cite{Buckley:2014ana} 6.2.1 are used.
} 
PDF sets with five active and massless flavours and their 
corresponding $\alphaS$ parametrisations with $\alphaS(m_Z)=0.118$ 
are used. 
While the use of the CT14nlo PDF set for NLO EW calculations is, 
strictly speaking, inconsistent due its missing QED evolution of 
the initial state quarks, it offers the quantification of the 
importance of photon initiated processes in these technical comparisons. 
To the same end, the CT14qed PDF set is not employed using its the best fit 
value for the intrinsic inelastic photon momentum fraction at the 
reference scale of $Q=1.295\,\text{GeV}$ 
$p_0^\gamma=0.05\%$, but rather evaluate it for the extremes of 
its $1\sigma$ uncertainty, $p_0^\gamma=0$ and $p_0^\gamma=0.14\%$, 
where applicable. 

In processes where jets need to be constructed to define 
a fiducial phase space volume, anti-$k_t$ jets with $R=0.4$ 
and $p_\perp>30\,\text{GeV}$ are used \cite{Cacciari:2008gp} 
and both partons and photons are considered as constituents. 
Similarly, if leptons need to be defined for the same purpose, 
they are dressed with all photons in a cone of $\Delta R=0.1$.
The number of phase space points in the computation of 
$\sigmaIRD(\{\alphadip\},\kappa,\cykt)$ is kept constant 
for each process, such that the indicated statistical uncertainty 
can be interpreted as a measure of the change of convergence 
of the subtraction with respect to the variation of the technical 
parameter choices.

\begin{figure}[t]
  \includegraphics[width=0.33\textwidth]{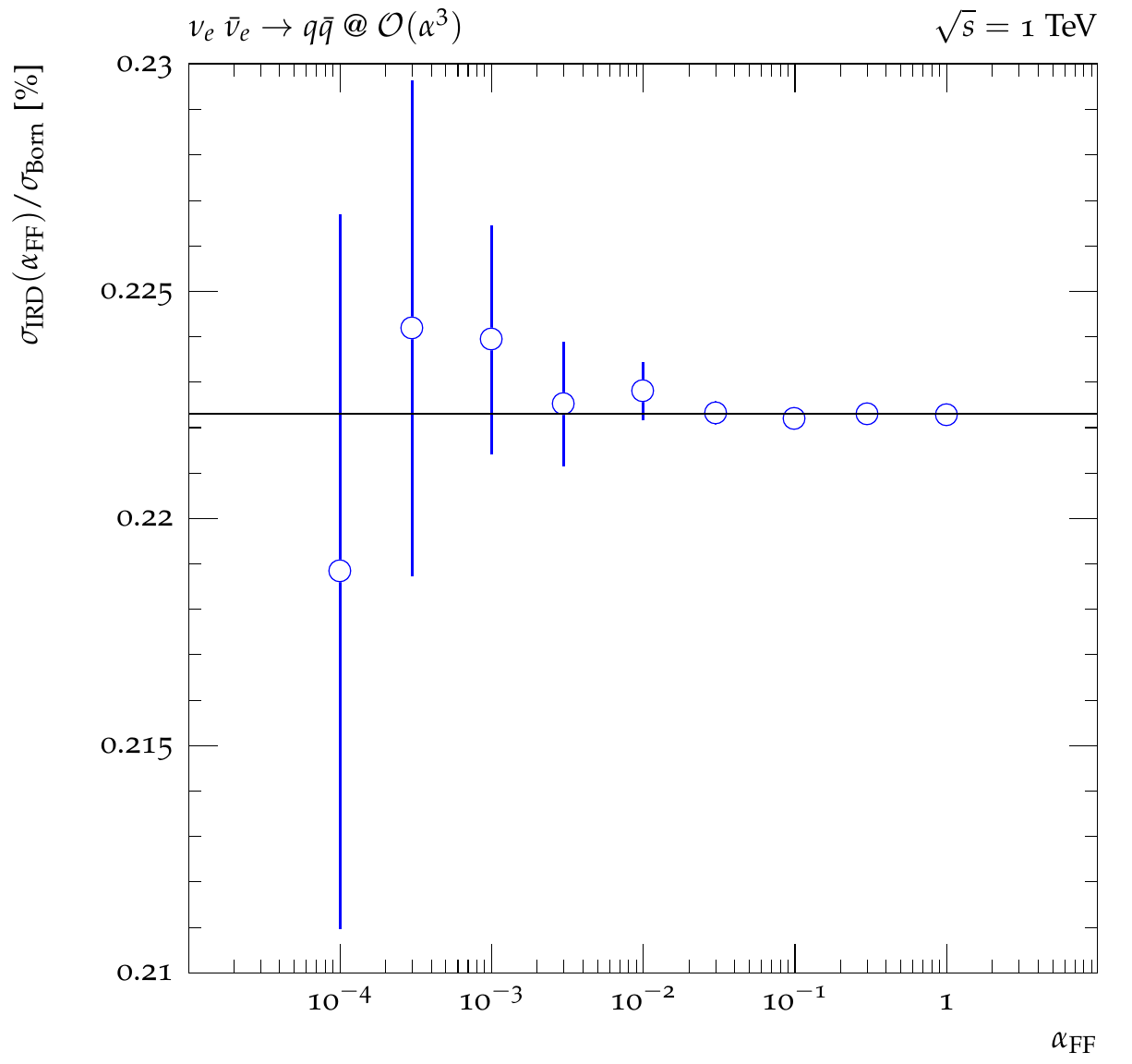}
  \includegraphics[width=0.33\textwidth]{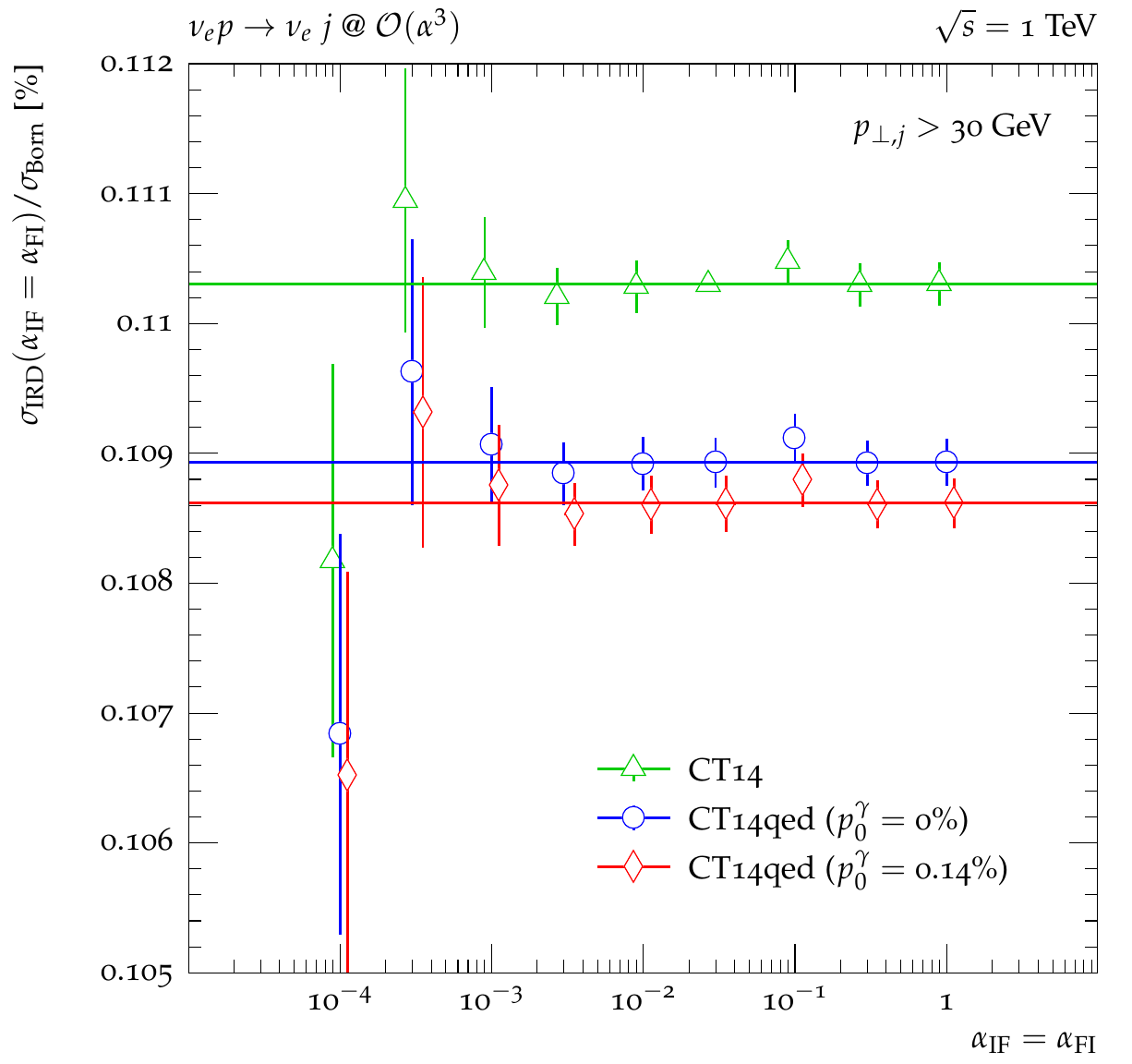}
  \includegraphics[width=0.33\textwidth]{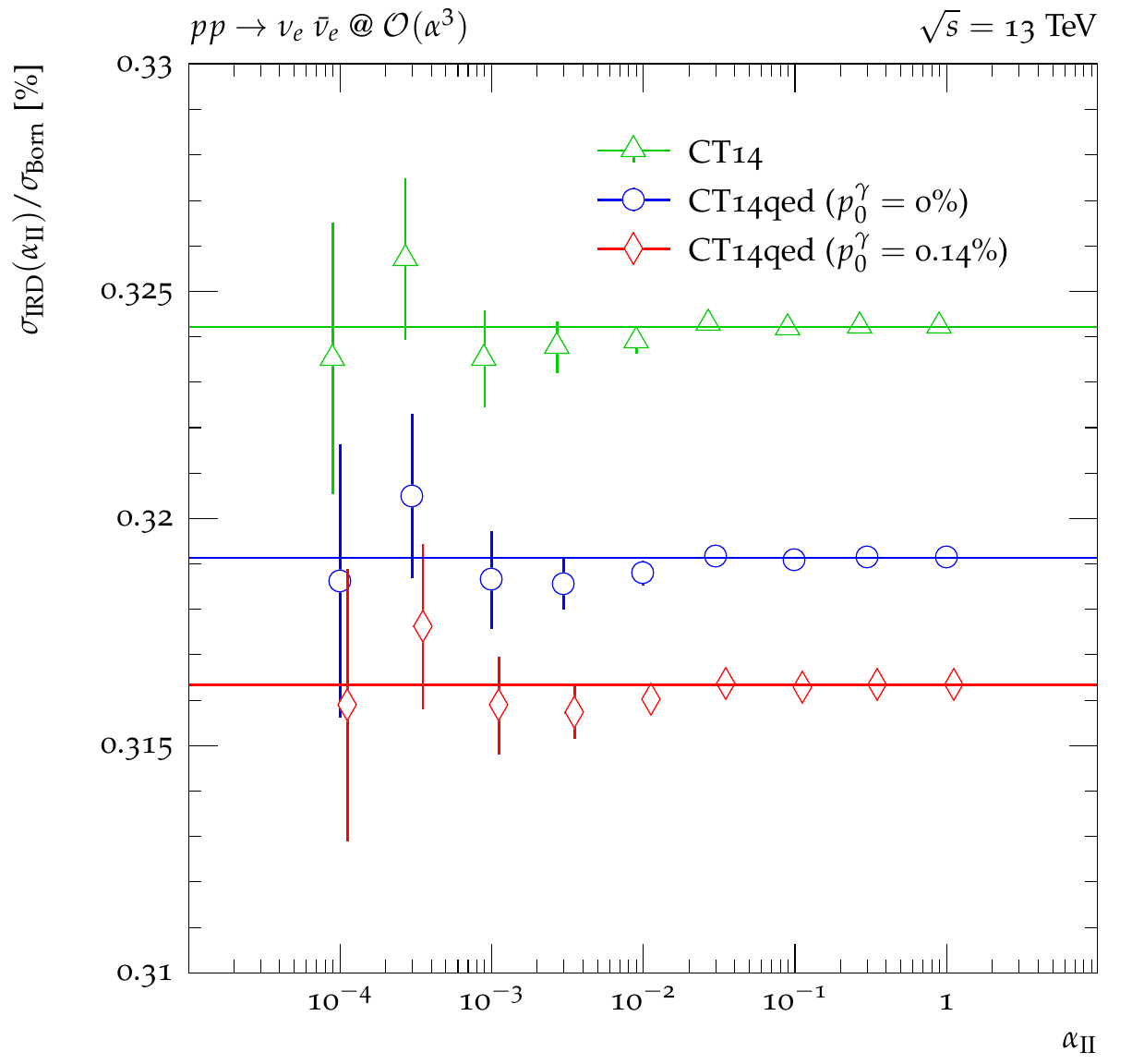}
  \caption{
	    $\{\alphadip\}$-dependence of the sum of 
	    integrated subtraction term and differentially subtracted 
	    real emission for $\nu_e\bar\nu_e\to jj$, $\nu_e p\to \nu_e j$ and 
	    $pp\to \nu_e\bar\nu_e$.
	    \label{fig:xs-alpha-nunujj-nupnuj-ppnunu}
          }
\end{figure}

\paragraph{Massless dipoles.}

Contrary to the QCD case, in the Standard Model almost all particles 
carry QED charges and therefore participate in the construction of dipoles. 
One notable exception are neutrinos. 
Therefore, in order to investigate the behaviour of the massless II, 
IF, FI and FF dipoles, and, thus, the independence of $\alphaII$, 
$\alphaIF$, $\alphaFI$ and $\alphaFF$ separately in this technical 
validation $\sigmaIRD$ is considered for all three different rotations 
of the interaction of a quark-anti-quark pair with a neutrino-anti-neutrino 
pair. 
Hence, besides the $\sigmaIRD$ contribution to the $\order(\alpha)$ 
correction to $pp\to \nu_e\bar{\nu}_e$ at the LHC at an invariant mass 
of 13\,TeV, $\sigmaIRD$ is computed for both the production of at least 
two jets at a hypothetical $\nu_e$-$\bar{\nu}_e$ collider at a 
centre-of-mass energy of 1\,TeV and inclusive single jet production in 
equally hypothetical $\nu_ep$ deep inelastic scattering (DIS) with the 
same centre-of-mass energy are calculated.

Figure \ref{fig:xs-alpha-nunujj-nupnuj-ppnunu} now details $\sigmaIRD$ 
for all three setups. 
$\nu_e\bar{\nu}_e\to jj$ production, detailed in the left hand side 
plot, comprises only FF dipoles and, thus, only depends on $\alphaFF$. 
Varying its value over four orders of magnitude leads leaves the 
value of $\sigmaIRD$ unchanged within the statistical uncertainties. 
The black line is placed at the central value of the computation with 
the smallest statistical uncertainty to guide the eye. 
As is evident, lowering the $\alpha$ parameter too much, i.e.\ 
restricting the subtraction to act only on configurations very close 
to the divergence, results in large cancellations between the 
real-subtracted and the integrated dipole contributions of eq.\ 
\eqref{eq:def-IRD}, degrading the statistical prowess of the 
calculation. 
Similarly, $pp\to\nu_e\bar{nu}_e$ production, detailed on the right 
hand side, comprises only II dipoles and, thus, depends on $\alphaII$ 
only. 
Showing the results for all three choices of photon content, contributing 
directly through the respective $\gamma q/\gamma\bar{q}$ channels as well 
as indirectly through the impact on the quark PDFs and, through momentum 
conservation, on the gluon PDF, a similar picture as for 
$\nu_e\bar{\nu}_e\to jj$ emerges. 
The three coloured lines now indicate the central value of the calculation 
with the smallest statistical uncertainty for each PDF choice. 
In each case, $\sigmaIRD$ is found to be stable under the variation 
of $\alphaII$.
Finally, the centre plot shows the correction contribution for the 
hypothetical DIS process $\nu_ep\to\nu_ej$ requiring at least one 
jet in the lab frame. 
As FI and IF dipoles always occur in pairs the $\alphaFI$ and $\alphaIF$ 
dependence is evaluated together here. 
The resulting $\sigmaIRD(\alphaIF=\alphaFI)$ are also found to be 
stable when varying both parameters simultaneously over four orders 
of magnitude.

\begin{figure}[t]
  \includegraphics[width=0.33\textwidth,angle=0,origin=c]{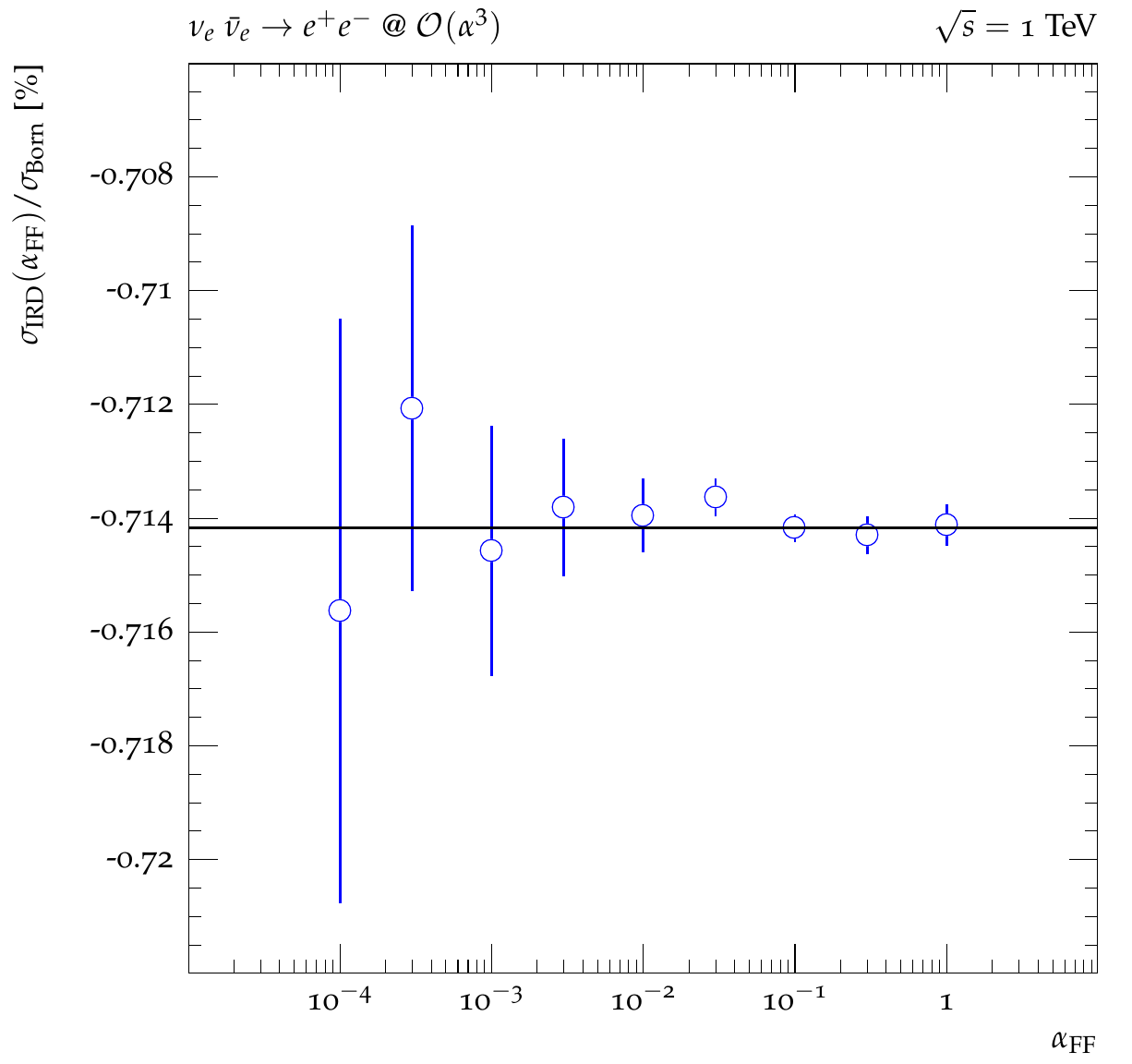}
  \includegraphics[width=0.33\textwidth,angle=0,origin=c]{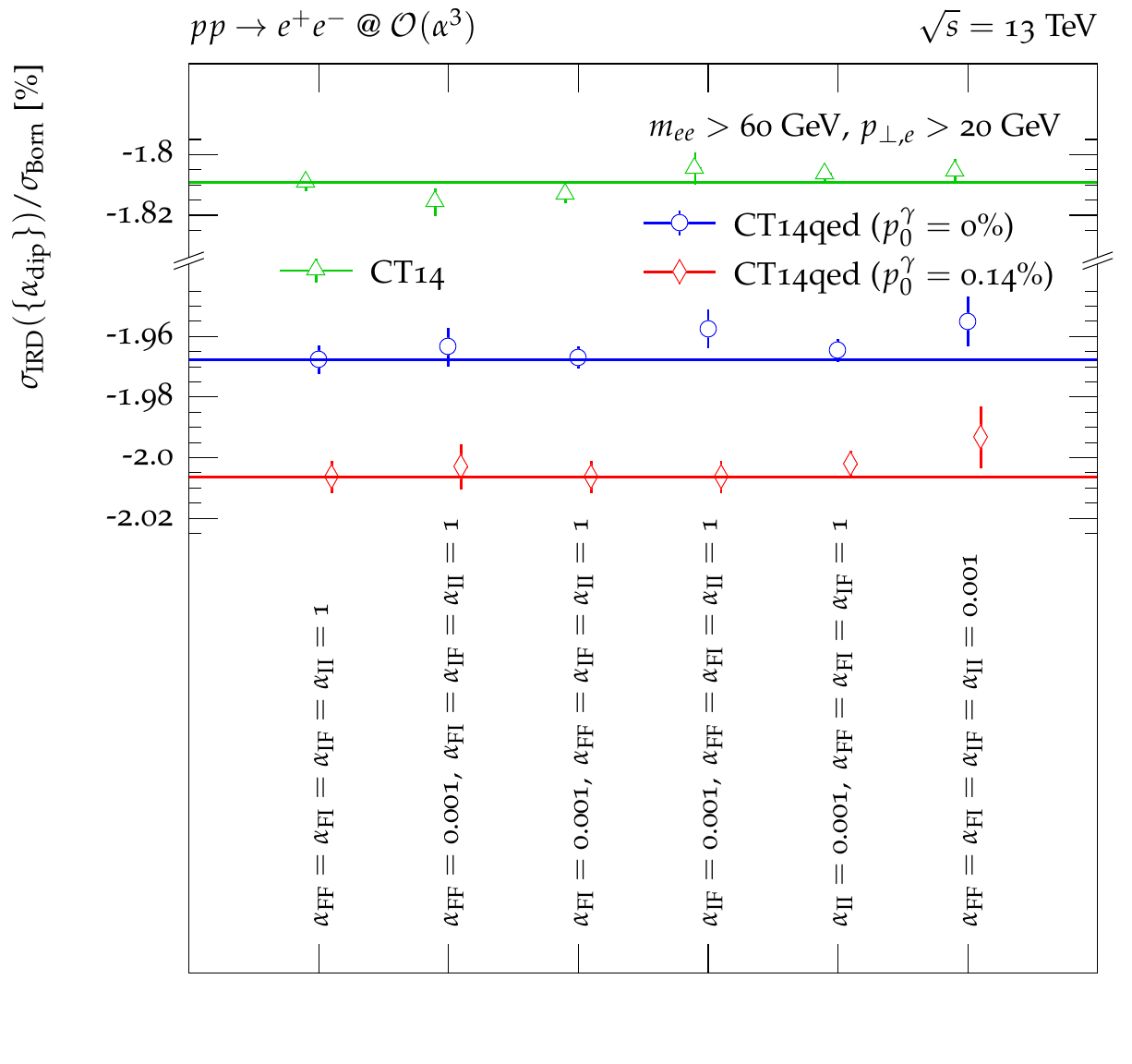}
  \includegraphics[width=0.33\textwidth,angle=0,origin=c]{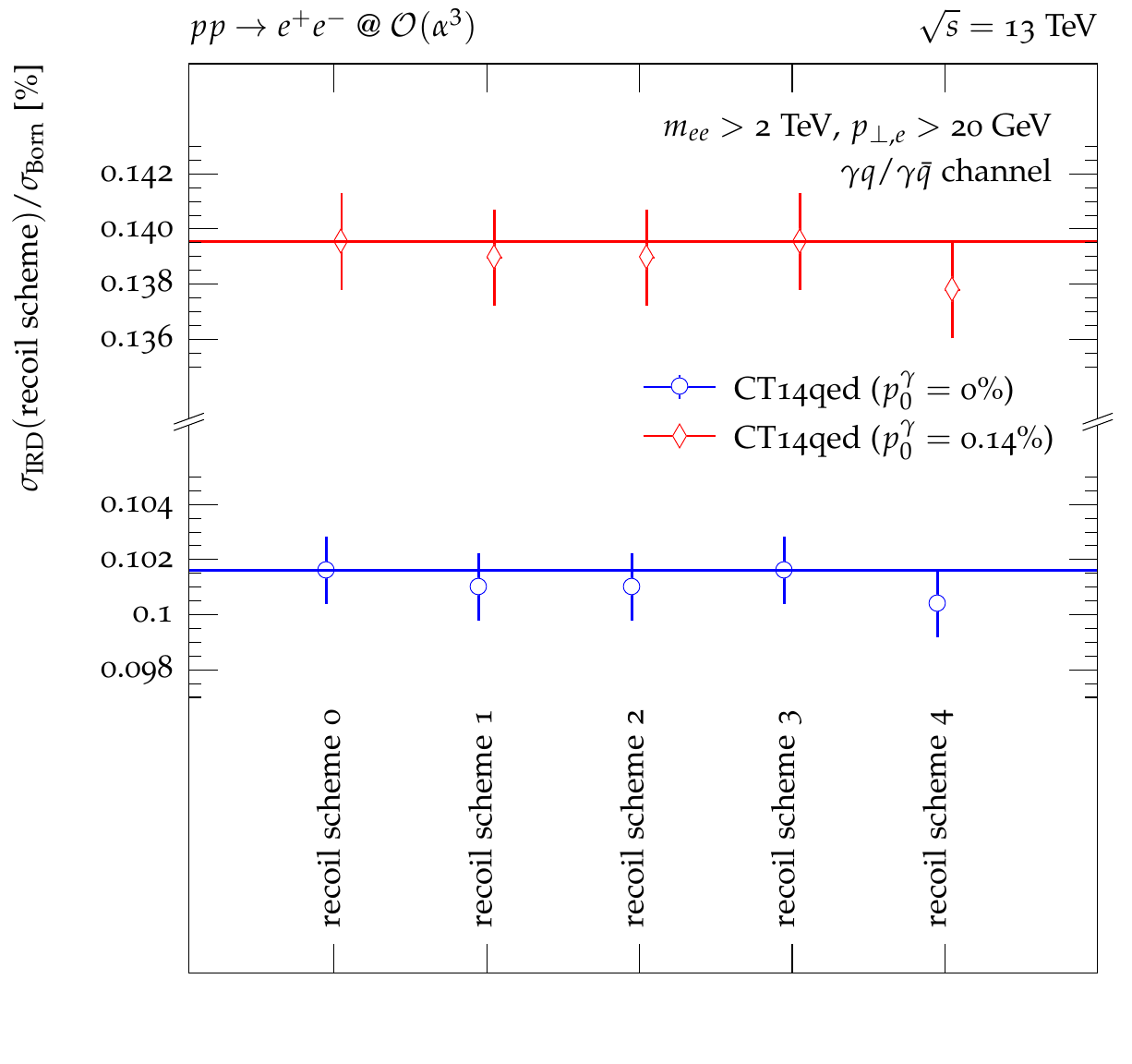}
  \caption{
	    \underline{Left and centre:} $\{\alphadip\}$-dependence of 
	    the sum of integrated subtraction term and differentially 
	    subtracted real emission for $\nu_e\bar\nu_e\to e^+e^-$ and 
	    $pp\to e^+e^-$. \\
	    \underline{\smash{Right:}} Dependence of the sum 
	    of the integrated subtraction term and differentially subtracted real emission 
	    for $pp\to e^+e^-$ in the $\gamma q$/$\gamma\bar{q}$ 
	    channel on the choice of recoil partner for the initial 
	    state photon splittings $\cykt$. 
	    The invariant mass of the electron pair is raised to 
	    increase the contribution of the photon induced channels. 
	    The $q\bar{q}$ and $\gamma\gamma$ channels comprise no 
	    dipoles with splitting photons.
	    \label{fig:xs-alpha-nunuee-ppee-ppeerecscheme}
          }
\end{figure}

Moving away from the simplest configurations, 
Figure \ref{fig:xs-alpha-nunuee-ppee-ppeerecscheme} displays the results 
for electron-positron pair production. 
The left hand side plot again displays their production at the hypothetical 
$\nu_e$-$\bar{\nu}_e$ collider used before, finding very similar results 
and their independence of $\alphaFF$. 
The centre plot now, however, displays the production of an electron-positron 
pair at the LHC. 
At leading order, this process proceeds through $q\bar{q}\to e^-e^+$ and 
$\gamma\gamma\to e^-e^+$ at $\order(\alpha^2)$. 
Consequently, the $q\bar{q}$ channel exhibits six dipoles of all four types. 
In the $\gamma\gamma$ channel, the number and types of dipoles present 
depends on the choice of possible photon splitting spectators $\cykt$. 
To regulate all LO singularities the fiducial phase space is defined 
by requiring the dressed electrons to have a transverse momentum of at 
least $20\,\text{GeV}$ and the pair to have an invariant mass of at 
least $60\,\text{GeV}$. 
As $\sigmaIRD$ now potentially depends on the full set $\{\alphadip\}$ 
no continuous variation over four orders of magnitude is performed. 
Instead, each of the four parameters is varied independently to 
$0.001$ keeping all others at their default value of $1$. 
These four variations are completed by setting 
$\alphaFF=\alphaFI=\alphaIF=\alphaII=1$ and $0.001$. 
The resulting correction contributions are found to be independent 
of $\{\alphadip\}$.

Finally, the right hand side plot of 
Figure \ref{fig:xs-alpha-nunuee-ppee-ppeerecscheme} displays the 
dependence of $\sigmaIRD$ on the choice of spectators in 
photon splittings. 
Only the $\gamma q/\gamma\bar{q}$ channel is considered as both 
the $\gamma\gamma$ and $q\bar{q}$ channels are independent of 
this choice in this process.
The $\gamma q/\gamma\bar{q}$ channel, however, still receives 
contributions from photon radiation off quarks in addition to 
the sought after photon splittings into quark-antiquark pairs. 
Hence, the minimum invariant mass is raised to $2\,\text{TeV}$ 
to increase the relative importance of the photon PDF, enhancing 
the photon splitting contribution. 
The resulting correction contribution at $\order(\alpha^3)$ is 
found to be independent of all five choices of photon splitting 
spectators available. 
Please note that for this process scheme 0 and 3 as well as 
scheme 1 and 2 lead to the same set of allowed spectators, 
respectively, and therefore to identical results.

\paragraph{Massive dipoles.}

\begin{figure}[t]
  \centering
  \includegraphics[width=0.33\textwidth,angle=0,origin=c]{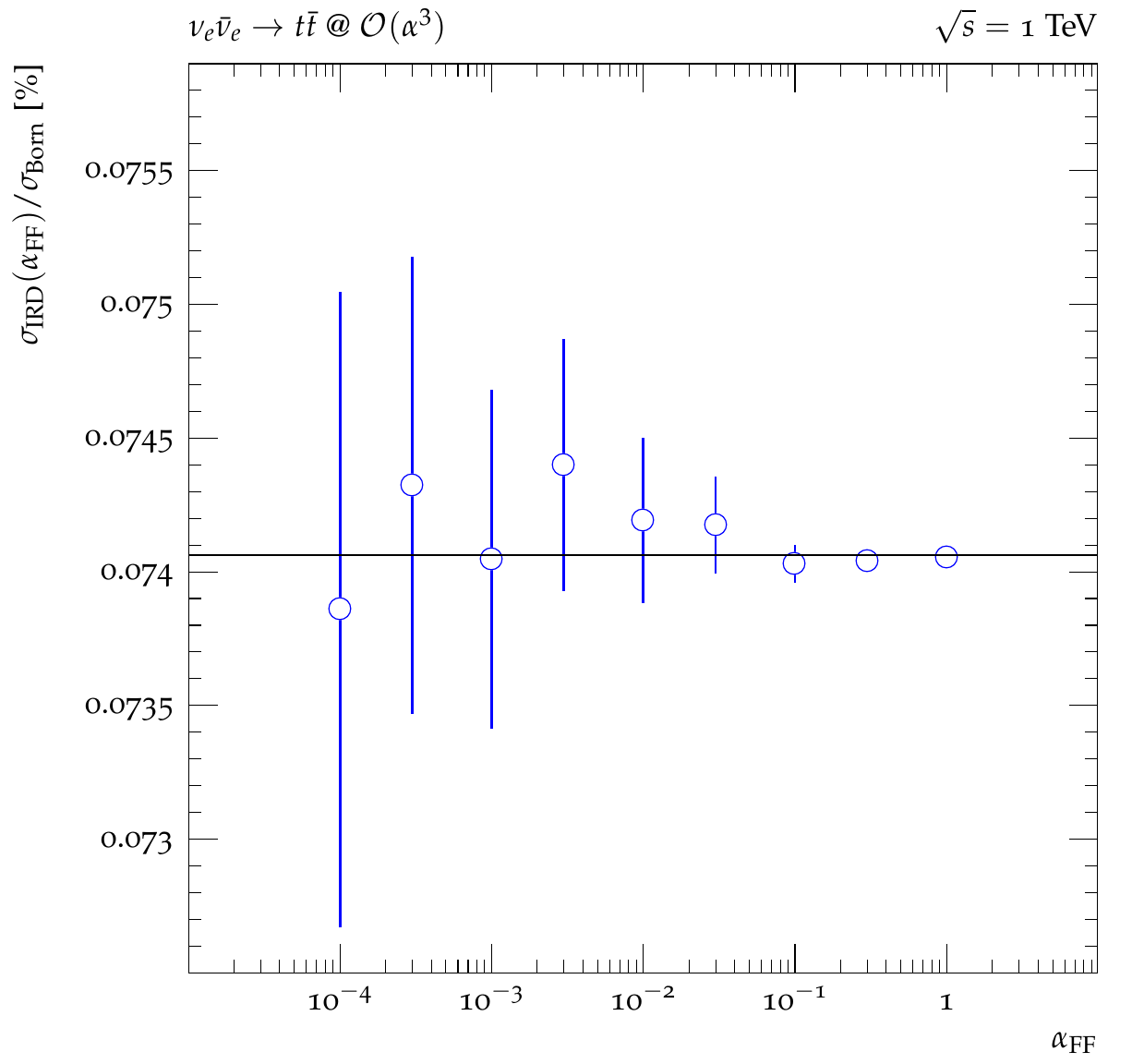}
  \hspace*{0.1\textwidth}
  \includegraphics[width=0.33\textwidth,angle=0,origin=c]{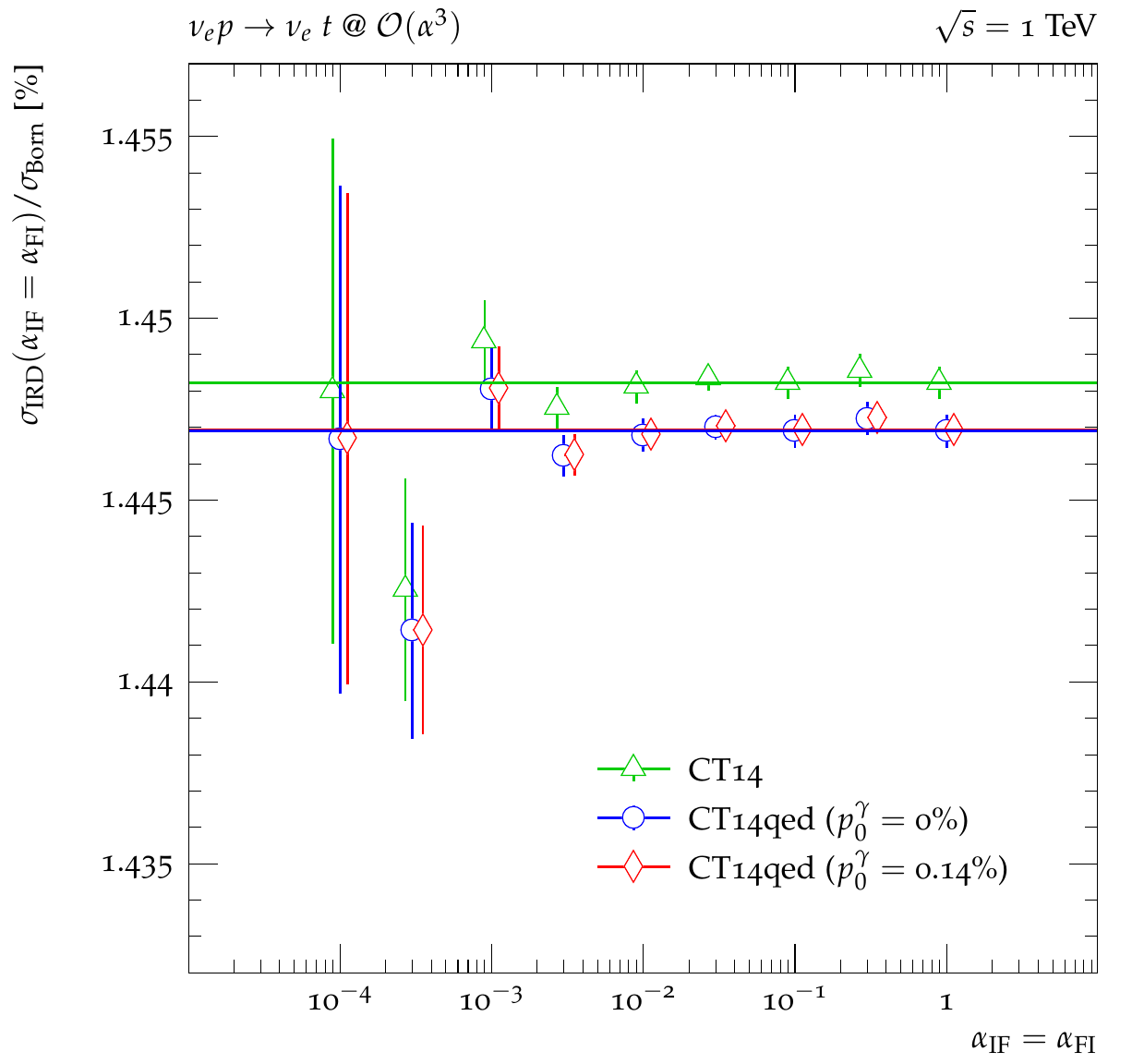}\\[2mm]
  \caption{
	    $\{\alphadip\}$-dependence of the sum of 
	    integrated subtraction term and differentially subtracted 
	    real emission for $\nu_e\bar\nu_e\to t\bar{t}$ and 
	    $\nu_e p\to\nu_e t$. 
	    For the latter the Standard Model is extended by a $u\bar{t}Z$ 
	    interaction with the structure and coupling as the existing 
	    $u\bar{u}Z$ interaction.
	    \label{fig:xs-alpha-nunutt-nupnut}
          }
\end{figure}

Massive particles are so far only allowed in the final state. 
\footnote{
  For advances for NLO calculations with initial state 
  massive particles see \cite{Dittmaier:1999mb,Basso:2015gca,Krauss:2017wmx}.
}
Dipoles involving massive partons, either as emitter or spectator, 
comprise only three types: FF, FI and IF.
Emittees are always considered massless, otherwise no singularity would 
be present.
In Figure \ref{fig:xs-alpha-nunutt-nupnut} again top-anti-top 
pair production at a hypothetical $\nu_e$-$\bar{\nu}_e$ collider and 
single top production at a hypothetical $\nu_e$-$p$ collider is 
considered in order to study the individual dipoles separately. 
In the left plot, the $\alphaFF$ (in)dependence of the 
$\order(\alpha^3)$ correction contribution $\sigmaIRD$, containing 
only massive FF dipoles, is shown. 
It exhibits the familiar picture of decreasing statistical prowess of 
the calculation with too small $\alphaFF$, but otherwise consistent 
results. 
The right plot details the $\alphaIF$ and $\alphaFI$ dependent 
massive dipoles in the hypothetical DIS scenario. 
As before, $\alphaIF$ and $\alphaFI$ are varied simultaneously 
for this purpose and the independence of the corrections contribution 
on both parameters is observed. 

\begin{figure}[t]
  \centering
  \includegraphics[width=0.33\textwidth,angle=0,origin=c]{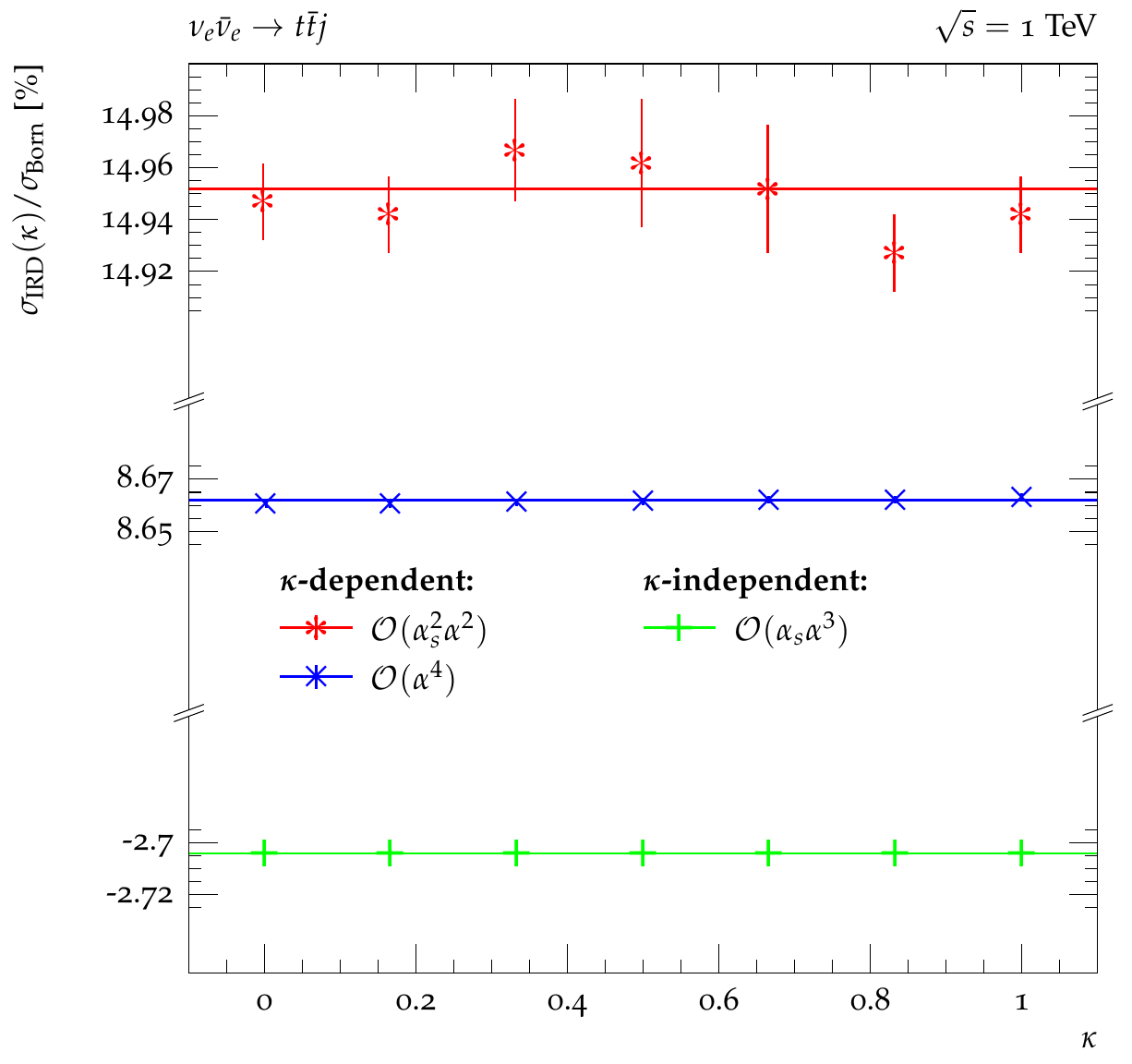}
  \hspace*{0.1\textwidth}
  \includegraphics[width=0.33\textwidth,angle=0,origin=c]{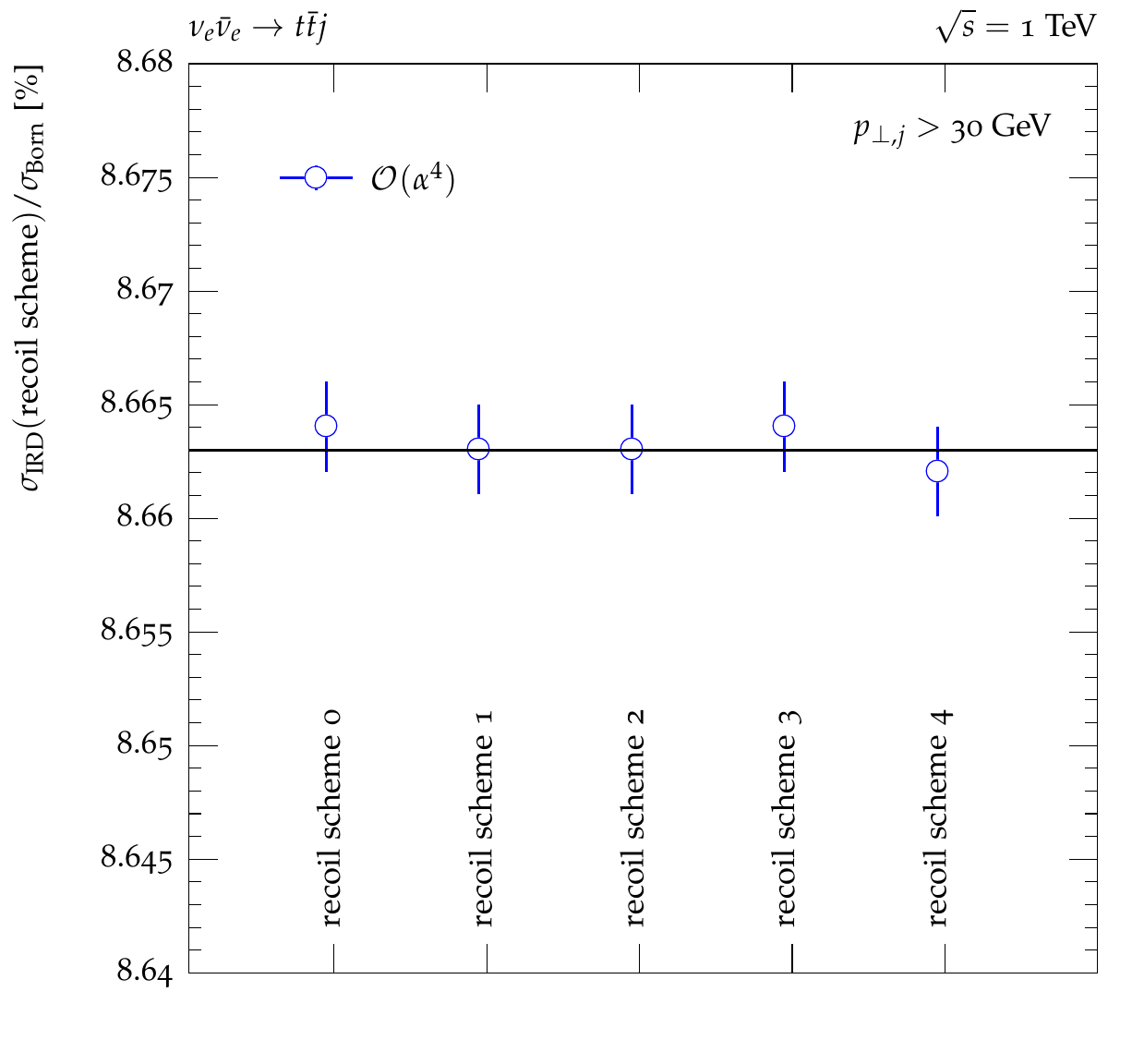}
  \caption{
	    \underline{\smash{Left:}} $\kappa$-dependence of the sum of
	    integrated subtraction term and differentially subtracted 
	    real emission for $\nu_e\bar\nu_e\to t\bar{t}j$.\\
	    \underline{\smash{Right:}} Dependence of the sum 
	    of the integrated subtraction term and differentially 
	    subtracted real emission for $\nu_e\bar\nu_e\to t\bar{t}j$ 
	    at $\order(\alpha^4)$ on the choice of recoil partner for the 
	    final state photon splittings $\cykt$.
	    \label{fig:xs-kappa-nunuttj}
          }
\end{figure}

The left plot of Figure \ref{fig:xs-kappa-nunuttj} 
now investigates the dependence of $\sigmaIRD$ 
on the $\kappa$ parameter. 
In only arises in FF dipoles of gluons splitting into gluons or massless 
quarks or photons splitting into massless fermions in the presence 
of a massive spectator.
Consequently, to restrict the number of additional contributions, 
top-anti-top pair production in association with one jet at the 
hypothetical $\nu_e$-$\bar{\nu}_e$ collider is considered. 
At LO, this process contributes is defined both at 
$\order(\alphaS\alpha^2)$ and $\order(\alpha^3)$ where the final 
state photon forms the jet. 
At NLO, there are contributions at $\order(\alphaS^2\alpha^2)$, 
$\order(\alphaS\alpha^3)$ and $\order(\alpha^4)$. 
The $\order(\alphaS\alpha^3)$ contribution, however, contains 
neither gluon nor photon splittings. 
Due to the relative size of $g\to gg$ and $g\to q\bar{q}$ splittings 
in relation to gluon radiation off the top quarks the $\kappa$ 
dependence of the $\order(\alphaS^2\alpha^2)$ contribution is much 
more pronounced than at $\order(\alpha^4)$, where photon radiation 
off the top quarks overwhelms the photon splitting contribution. 
Nonetheless, at both orders an independence of the $\sigmaIRD$ of 
$\kappa$ is found. 
In addition, the right plot investigates the influence on the 
choice of spectators for the final state photon splitting 
ocurring at $\order(\alpha^4)$. 
No dependence on this choice is observed. 
Please note that for this process scheme 0 and 3 as well as 
scheme 1 and 2 lead to the same set of allowed spectators, 
respectively, and therefore to identical results.

\begin{figure}[t!]
  \centering
  \includegraphics[width=0.33\textwidth,angle=0,origin=c]{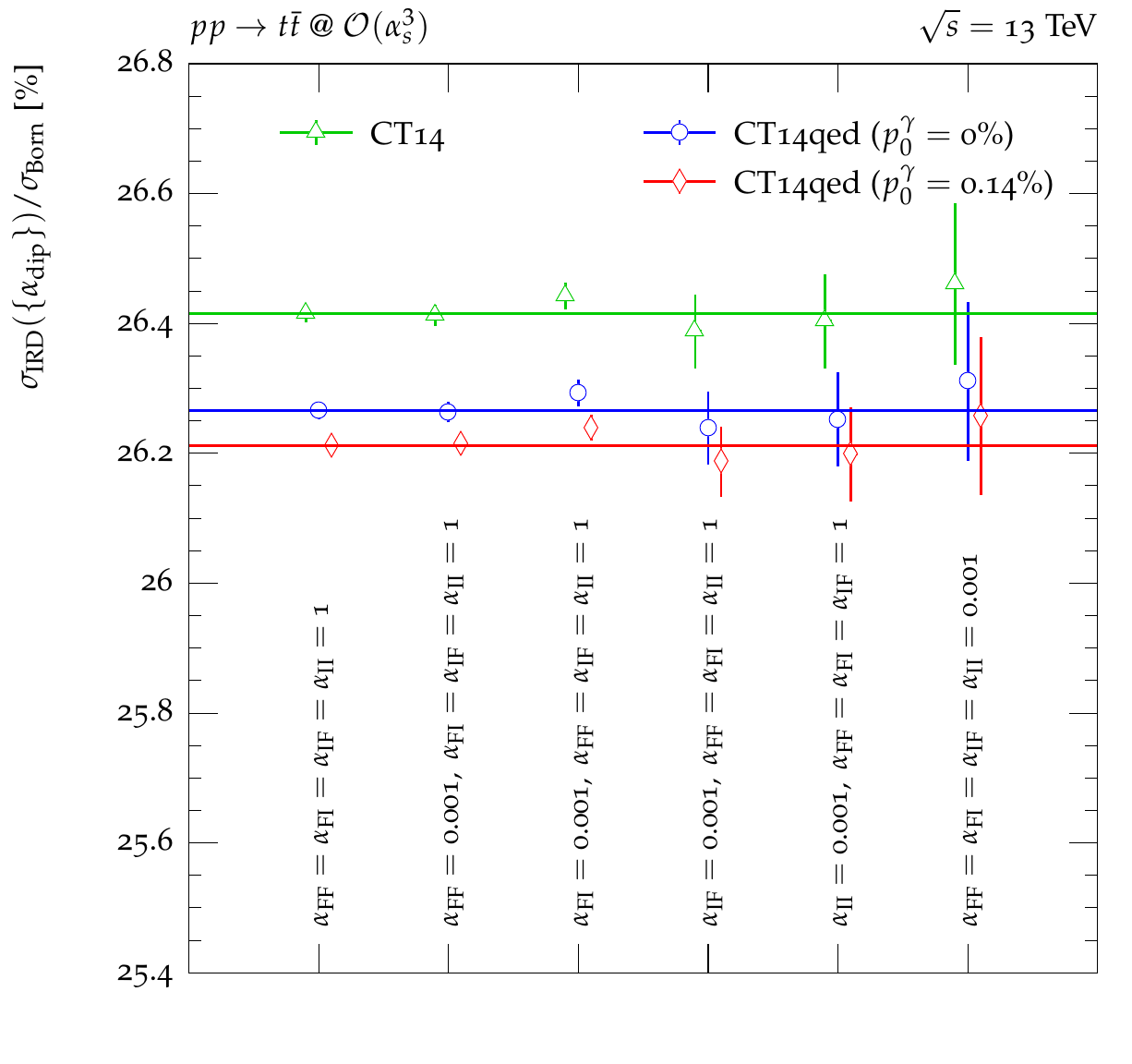}
  \hspace*{0.1\textwidth}
  \includegraphics[width=0.33\textwidth,angle=0,origin=c]{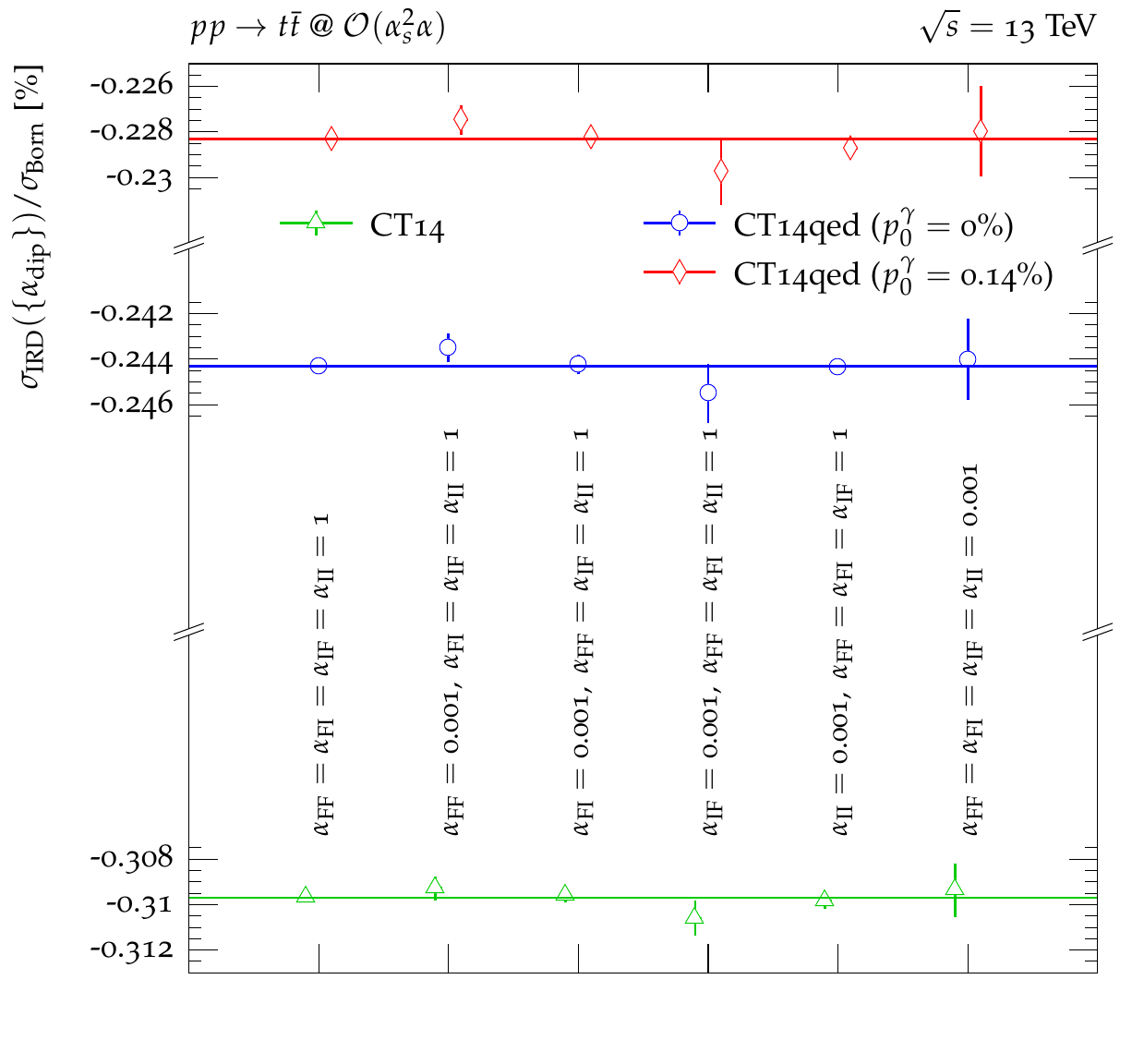}\\
  \includegraphics[width=0.33\textwidth,angle=0,origin=c]{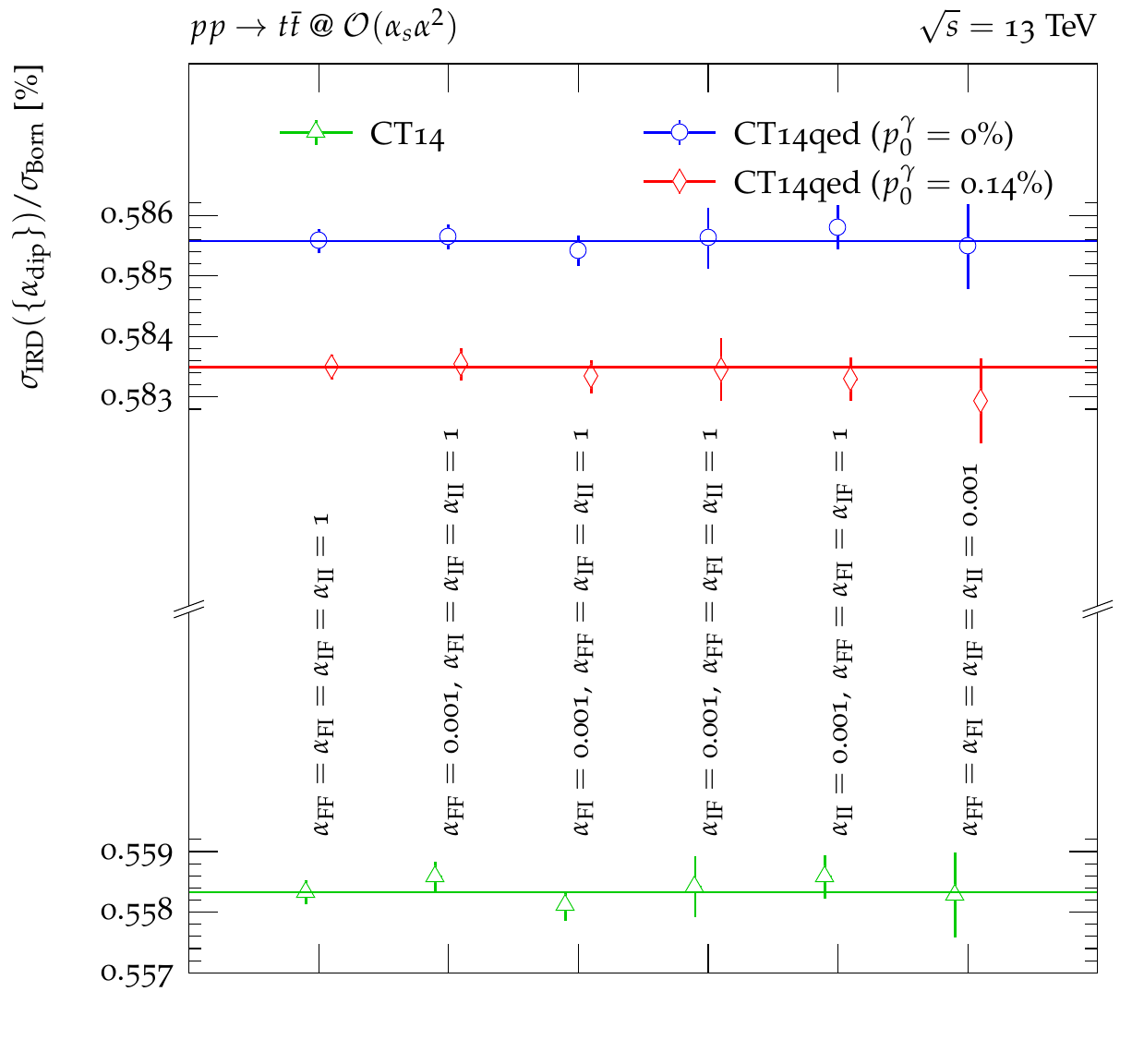}
  \hspace*{0.1\textwidth}
  \includegraphics[width=0.33\textwidth,angle=0,origin=c]{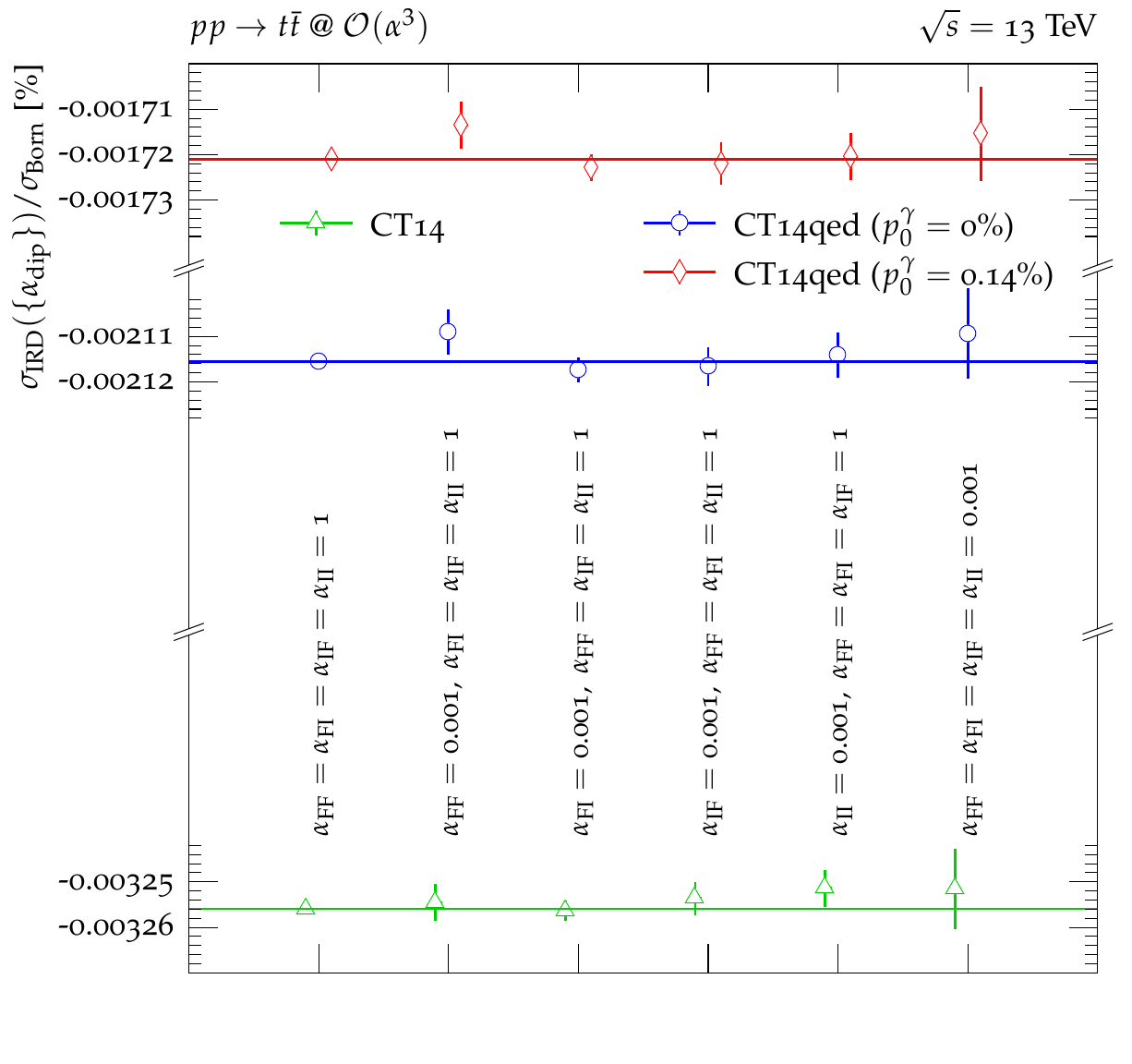}
  \caption{
	    $\{\alphadip\}$-dependence of the sum of 
	    integrated subtraction term and differentially subtracted 
	    real emission for $\nu_e\bar\nu_e\to t\bar{t}$ and 
	    $pp\to t\bar{t}$.
	    \label{fig:xs-alpha-pptt}
          }
\end{figure}

Figure \ref{fig:xs-alpha-pptt} now considers top-anti-top pair production 
at the LHC. 
This process occurs at LO at $\order(\alphaS^2)$, 
$\order(\alphaS\alpha)$ and $\order(\alpha^2)$. 
Thus, at NLO there exist four contributions, at $\order(\alphaS^3)$, 
$\order(\alphaS^2\alpha)$, $\order(\alphaS\alpha^2)$ and $\order(\alpha^3)$. 
While the $\order(\alphaS^3)$ and $\order(\alpha^3)$ terms 
can be clearly identified as NLO QCD and NLO EW corrections to 
the LO $\order(\alphaS^2)$ and $\order(\alpha^2)$ expressions, 
respectively, the $\order(\alphaS^2\alpha)$ and $\order(\alphaS\alpha^2)$ 
terms do not possess such a unique characterisation: they act as both 
NLO QCD corrections and NLO EW corrections to different LO 
processes. 
Consequently, their divergence structure contains singularities of both 
QCD and QED origin. 
Hence, both QCD and QED dipoles with underlying Born processes of 
different orders are needed for a full subtraction of all divergences. 
As explained in Section \ref{sec:builddip}, this holds true both 
for the differential and integrated subtraction terms. 
It therefore serves as an additional check to verify the independence 
of the result of the $\{\alphadip\}$ for each $\order(\alphaS^{3-m}\alpha^m)$, 
$m=0..3$, individually. 
And indeed, the correction contribution $\sigmaIRD$ is found independent 
of $\{\alphadip\}$ for each such order.

\paragraph{External $\boldsymbol W$ bosons.}

\begin{figure}[t!]
  \centering
  \includegraphics[width=0.33\textwidth]{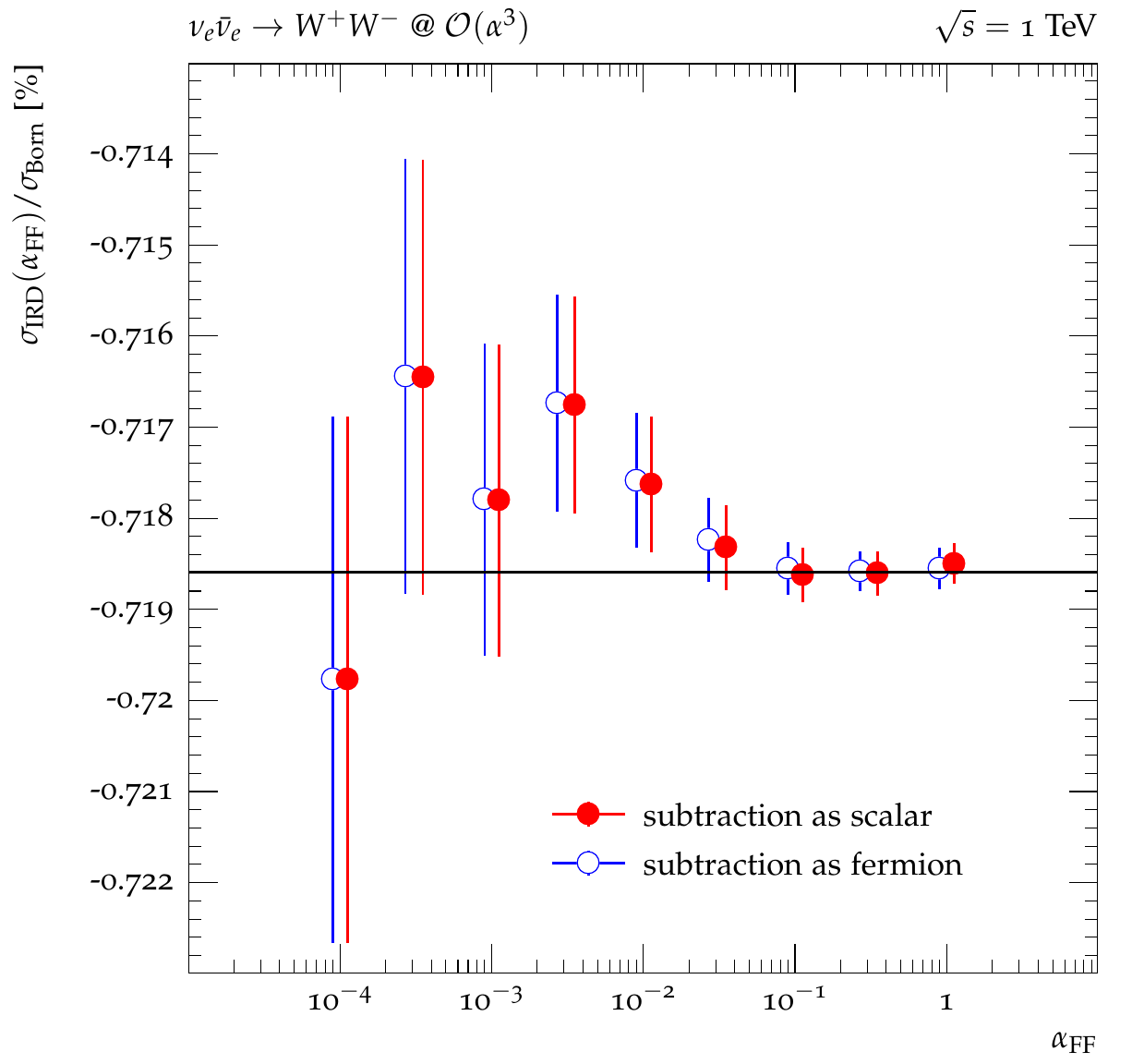}
  \hspace*{0.1\textwidth}
  \includegraphics[width=0.33\textwidth]{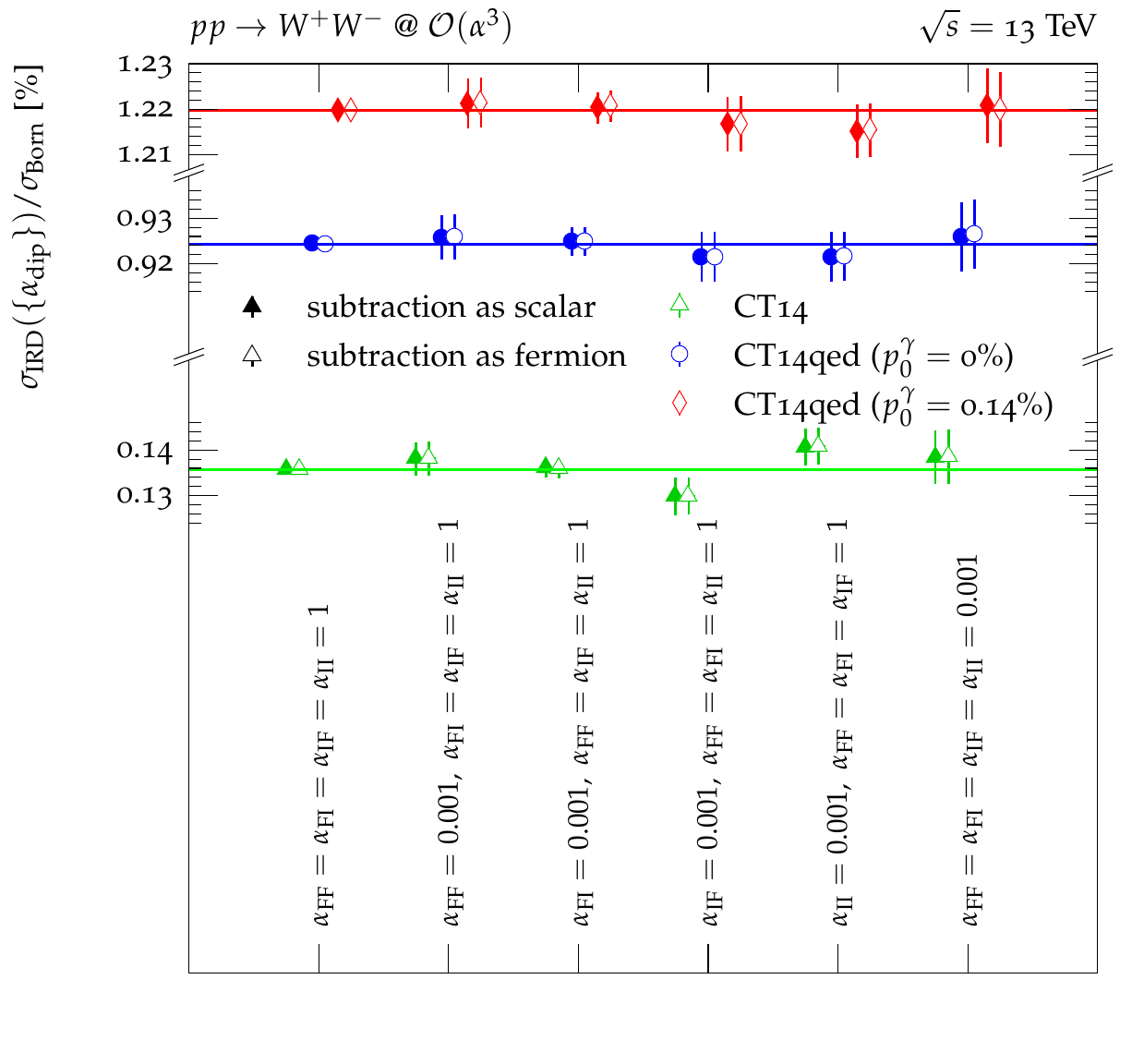}
  \caption{
	    $\{\alphadip\}$-dependence of the sum of 
	    integrated subtraction term and differentially subtracted 
	    real emission for $\nu_e\bar\nu_e\to W^+W^-$ and $pp\to W^+W^-$. 
	    Both choices of subtraction terms for external massive charged 
	    vector bosons, using the ones of an external massive scalar 
	    (filled symbols) and massive fermions (empty symbols), are compared.
	    \label{fig:xs-alpha-nunuWW-ppWW}
          }
\end{figure}

Lastly, it may become necessary to also consider external $W$ bosons 
(or other massive charged particle with spin $>\tfrac{1}{2}$ in 
BSM theories) as stable final state particles, e.g.\ to reduce the 
computational complexity for high final state multiplicity processes 
where off-shell effects and effects in the decays can be ignored 
or recovered through other means \cite{Schonherr:2008av}.
In this case the literature does not provide expressions for the 
respective massive dipole functions. 
As their mass, however, can be assumed to be large enough to suppress 
collinear radiation well enough, this only leaves the spin-independent 
soft photon emission limit. 
\footnote{
  This assumption also proves to be correct for on-shell 
  $W$ production at a 100\,TeV collider.
}
Here, both the expressions for the radiation of a photon off a 
massive scalar or a massive fermion can be used.

Figure \ref{fig:xs-alpha-nunuWW-ppWW} details the production 
of a $W^+W^-$ pair in the hypothetical $\nu_e$-$\bar{\nu}_e$ 
collider, separating the FF dipoles and their $\alphaFF$ 
dependence, on the left hand side. 
As before, $\sigmaIRD$ is found to be independent within the 
statistical accuracy and also independent of the whether the 
massive scalar or massive fermion subtraction terms are used.
The right hand side focusses on their production at the LHC. 
Again, all dipoles contribute at $\order(\alpha^3)$, leading 
again to the afore described six-point variation. 
Also in this case, the result is independent of $\{\alphadip\}$ 
and the choice of scalar or fermionic subtraction term. 
The FI and IF dipoles cannot be investigated separately, as in 
the $\nu_ep\to\nu_et$ case, due to charge conservation.

%% file: text/conclusions.tex
\section{Conclusions}
\label{sec:conclusions}

This paper detailed the construction and implementation of 
the adaptation of the Catani-Seymour subtraction formalism 
for NLO EW calculations. 
Besides the translation of the QCD dipole functions to 
the QED case, several other issues have been addressed. 
They include the special role photon splittings play in 
the formalism, embedding extermal massive emitters of 
spin $>\tfrac{1}{2}$ into the formalism and the interplay 
of QCD and QED subtractions for processes exhibiting both 
kinds of divergences. 
The resulting general subtraction for NLO EW calculations 
has been implemented in the \Sherpa Monte-Carlo event 
generator framework. 
Interfaces to \OpenLoops, \GoSam and \Recola to access the 
needed virtual corrections exist and are fully functional.

In addition to the checks against independent implementations 
on the level of partial and total cross sections performed 
in previous publications, numerous internal cross checks for 
independence of technical parameter choices, 
$\{\alphadip\}=\{\alphaFF,\alphaFI,\alphaIF,\alphaII\}$, 
$\kappa$ and the choice of spectator in photon splittings $\cykt$,
have been presented here. 
This implementation will become publically available in the 
near future with the next major \Sherpa release 
and an extension to the \Comix matrix element 
generator \cite{Gleisberg:2008fv} is foreseen.

\subsection*{Acknowledgements}

MS would like to thank N.\ Greiner, S.\ H\"oche, S.\ Kallweit, J.\ Lindert, and S.\ Schumann 
for numereous checks of the implementation and S.\ Pozzorini for many 
useful discussion. MS acknowledges support by the Swiss National Foundation 
(SNF) under contract PP00P2–12855 as well as the European Union's Horizon 
2020 research and innovation programme as part of the Marie 
Skłodowska-Curie Innovative Training Network MCnetITN3 (grant agreement 
no.\ 722104) and the ERC Advanced Grant MC@NNLO (340983). 
MS would also like to thank the University Duisburg–Essen for the 
kind hospitality during essential parts of the project.
All plots are based on \Rivet's \cite{Buckley:2010ar} plotting tools.

%% file: text/appendix.tex
\section{Differential splitting functions}
\label{app:diffsplit}

This appendix details the complete functional form of the dipoles 
introduced in eq.\ \eqref{eq:dipole_factorisation}, 
$\mc{D}_{ij,k}$, $\mc{D}_{ij}^a$, $\mc{D}_{j,k}^{a}$ and $\mc{D}_j^{a,b}$. 
The functional forms of the spin-dependent insertions $\mf{V}_{ij,k}$, 
$\mf{V}_{ij}^a$, $\mf{V}_{j,k}^a$ and $\mf{V}_j^{a,b}$ 
and the kinematic maps of the differential splitting functions are 
taken from the original QCD case \cite{Catani:1996vz,Catani:2002hc}. 
They are summarised in the following.

\subsection{Final-final dipoles} 

Dipoles with both emitter and spectator in the final state take 
the form
\begin{equation}
  \begin{split}
    \mc{D}_{ij,k}
    \,=\;& -\frac{1}{(p_i+p_j)^2-m_\ijt^2}\;\Qop{\ijt\kt}\;
	   {}_m\langle\ldots,\ijt,\ldots,\kt,\ldots|\mf{V}_{ij,k}|
	              \ldots,\ijt,\ldots,\kt,\ldots\rangle{}_m\;.
  \end{split}
\end{equation}
Therein, the charge-correlator is defined in eq.\ \eqref{eq:chargecorrelator}.
All momenta of the dipole are on-shell
\begin{equation}
  \begin{split}
    p_i^2=&\;m_i^2\,, \qquad p_j^2=m_j^2\,,\qquad p_\ijt^2=m_\ijt^2\,,\qquad p_k^2=p_\kt^2=m_k\,,
  \end{split}
\end{equation}
and the total four momentum flowing through it is given by
\begin{equation}
  \begin{split}
    q=&\;p_i+p_j+p_k=p_\ijt+p_\kt\;.
  \end{split}
\end{equation}
It is thus invariant under the emission. 
The momenta of the parton before and after the splitting are connected 
through the map 
\begin{equation}
  \begin{split}
    p_\kt=\;&\sqrt{\frac{\Kallen{q^2}{m_\ijt^2}{m_\kt^2}}
                        {\Kallen{q^2}{(p_i+p_j)^2}{m_k^2\vphantom{m_\kt^2}}}}
             \left(p_k-\frac{qp_k}{q^2}\,q\right)
	     +\tfrac{1}{2}\,\frac{q^2+m_\kt^2-m_\ijt^2}{q^2}\,q\\
    p_\ijt=\;&q-p_\kt
  \end{split}
\end{equation}
wherein $m_\kt=m_k$ and the Kallen function is defined as 
$\Kallen{a}{b}{c}=a^2+b^2+c^2-2ab-2ac-2bc$\,. 
The splitting variables read
\begin{equation}
  \begin{split}
    y_{ij,k}=\frac{p_ip_j}{p_ip_j+p_ip_k+p_jp_k}
    \;,\qquad
    z_i=1-z_j=\frac{p_ip_k}{(p_i+p_j)p_k}
    \qquad\text{and}\qquad
    z_{i,j}^{(m)}=z_{i,j}-\frac{1-v_{ij,k}}{2}\;,
  \end{split}
\end{equation}
which are defined on the intervals $y_{ij,k}\in[y_-,y_+]$ and $z_i\in[z_-,z_+]$. 
The boundaries are given by
\begin{equation}
  \label{eq:ff-phasespaceboundaries}
  \begin{split}
    \begin{array}{ll}
      \displaystyle
      y_-
      \,=\;
	\frac{2m_im_j}{q_{ij,k}^2}
      \hspace*{40mm}
      &
      \displaystyle
      z_-
      \,=\;
	\frac{2m_i^2-q_{ij,k}^2y_{ij,k}}{2(m_i^2+m_j^2+q_{ij,k}^2y_{ij,k})}\,
	\left(1-v_{ij,i}v_{ij,k}\right)\\[5mm]
      \displaystyle
      y_+
      \,=\;
	1-\frac{2m_k\sqrt{q^2-m_k^2}}{q_{ij,k}^2}
      &
      z_+
      \displaystyle
      \,=\;
	\frac{2m_i^2-q_{ij,k}^2y_{ij,k}}{2(m_i^2+m_j^2+q_{ij,k}^2y_{ij,k})}\,
	\left(1+v_{ij,i}v_{ij,k}\right)\;.
    \end{array}
  \end{split}
\end{equation}
with $q_{ij,k}^2=q^2-m_i^2-m_j^2-m_k^2$. 
As can be seen, a divergence, residing at $y_{ij,k}=0$, is only present 
if either $i$ or $j$ are massless.
The relative velocities between $\ijt$ and $\kt$, $i+j$ and $k$, and 
$i+j$ and $i$ are given by
\begin{equation}
  \begin{split}
    v_{\ijt,\kt}
    \,=\;&
      \frac{\sqrt{\Kallen{q^2}{m_\ijt^2}{m_\kt^2}}}{q^2-m_\ijt^2-m_\kt^2}\\
    v_{ij,k}
    \,=\;&
      \frac{\sqrt{\left[2m_k^2+q_{ij,k}^2(1-y_{ij,k})\right]^2-4q^2m_k^2}}
           {q_{ij,k}^2(1-y_{ij,k})}
    \qquad\text{and}\qquad
    v_{ij,i}
    \,=\;&
      \frac{\sqrt{q_{ij,k}^4y_{ij,k}^2-4m_i^2m_j^2}}
           {q_{ij,k}^2y_{ij,k}+2m_i^2}
    \;.
  \end{split}
\end{equation}
They are introduced to facilitate the analytic integration and only 
take effect away from the singular limit. 
Thus, denoting the Spins of parton $\ijt$ in 
$\langle\ldots,\ijt,\ldots|\mf{V}_{ij,k}|\ldots,\ijt,\ldots\rangle$ 
by $s$, $s'$ (if $\ijt$ is a fermion), $\mu$, $\nu$ 
(if $\ijt$ is a photon), or omitting them (if $\ijt$ is a scalar) 
the dipole functions are defined as 
\begin{equation}
  \begin{split}
    \langle s|\mf{V}_{\gamma f,k}|s'\rangle
    \,=\;&
      8\pi\mu^{2\epsilon}\alpha
      \left\{
	\frac{2}{1-z_j(1-y_{ij,k})}
	-\frac{\tilde{v}_{ij,k}}{v_{ij,k}}
	\left[
	  2-z_i(1-\epsilon)+\frac{m_f^2}{p_ip_k}
	\right]
      \right\}
      \delta_{ss'}\\
    \langle\mu|\mf{V}_{f\bar{f},k}|\nu\rangle
    \,=\;&
      8\pi\mu^{2\epsilon}\alpha\,
      \frac{1}{v_{ij,k}}
      \left\{
	-g^{\mu\nu}
	\left[
	  1-\frac{2\kappa}{1-\epsilon}\left(z_+z_--\frac{m_f^2}{(p_i+p_j)^2}\right)
	\right]
      \right.\\
    & \hspace*{20.5mm}\left.{}
	-\frac{4}{(p_i+p_j)^2\vphantom{m_f^2}}
	 \left[z_i^{(m)}p_i^\mu-z_j^{(m)}p_j^\mu\right]
	 \left[z_i^{(m)}p_i^\nu-z_j^{(m)}p_j^\nu\right]
      \right\}\\
    \langle |\mf{V}_{\gamma s,k}|\rangle
    \,=\;&
      8\pi\mu^{2\epsilon}\alpha
      \left\{
	\frac{2}{1-z_j(1-y_{ij,k})}
	-\frac{\tilde{v}_{ij,k}}{v_{ij,k}}
	\left[
	  2+\frac{m_s^2}{p_ip_k}
	\right]
      \right\}\;.
  \end{split}
\end{equation}
It is thus clear that only in the case of photons splitting into 
massless fermions does the dipole splitting function have a 
non-diagonal structure. 
In all other cases, the insertion 
$\langle\ldots,\ijt,\ldots|\mf{V}_{ij,k}|\ldots,\ijt,\ldots\rangle$ 
reverts to a simple multiplication of a real-valued splitting 
function and a standard Born matrix element.
The parameter $\kappa$ controls finite terms and only takes effects 
in the case where the respective fermionic products of the splittings 
are massive. 
Setting $\kappa=0$ somewhat simplifies the spin-dependence of the 
differential dipoles.

\subsection{Final-initial dipoles}

Dipoles with the emitter in the final state and the spectator in the 
initial state take the form 
\begin{equation}
  \begin{split}
    \mc{D}_{ij}^a
    \,=\;& -\frac{1}{(p_i+p_j)^2-m_\ijt^2}\,\frac{1}{x_{ij,a}}\;\Qop{\ijt\at}\;
	   {}_m\langle\ldots,\ijt,\ldots,\at,\ldots|\mf{V}_{ij}^a|
	              \ldots,\ijt,\ldots,\at,\ldots\rangle{}_m
  \end{split}
\end{equation}
Therein, the charge-correlator is defined in eq. \eqref{eq:chargecorrelator}. 
All momenta of the dipole are on-shell
\begin{equation}
  \begin{split}
    p_i^2=&\;m_i^2\,, \qquad p_j^2=m_j^2\,,\qquad p_\ijt^2=m_\ijt^2\,,\qquad p_a^2=p_\at^2=0,
  \end{split}
\end{equation}
and the total four momentum flowing through it is given by
\begin{equation}
  \begin{split}
    q=&\;p_i+p_j-p_a=p_\ijt-p_\at\;.
  \end{split}
\end{equation}
It is thus invariant under the emission. The momenta of the parton 
before and after the splitting are connected through the map 
\begin{equation}
  \begin{split}
    p_\at=\;&x_{ij,a}p_a\\
    p_\ijt=\;&q+p_\at\;.
  \end{split}
\end{equation}
The splitting variables read
\begin{equation}
  \begin{split}
    x_{ij,a}=1-\frac{p_ip_j-\tfrac{1}{2}\left(m_\ijt^2-m_i^2-m_j^2\right)}{(p_i+p_j)p_a}
    \qquad\text{and}\qquad
    z_i=1-z_j=\frac{p_ip_a}{(p_i+p_j)p_a}\;.
  \end{split}
\end{equation}
The singularity of the splitting resides at $x_{ij,a}=1$ and is only 
present if either $m_\ijt=m_i$ or $m_\ijt=m_j$. 
Adopting the above convention for labeling the emitter's spins
the dipole functions are defined as 
\begin{equation}
  \begin{split}
    \langle s|\mf{V}_{\gamma f}^a|s'\rangle
    \,=\;&
      8\pi\mu^{2\epsilon}\alpha
      \left\{
	\frac{2}{2-x_{ij,a}-z_j}
	-2+z_i(1-\epsilon)-\frac{m_f^2}{p_ip_j}
      \right\}
      \delta_{ss'}\\
    \langle\mu|\mf{V}_{f\bar{f}}^a|\nu\rangle
    \,=\;&
      8\pi\mu^{2\epsilon}\alpha
      \left\{
	-g^{\mu\nu}-\frac{4}{(p_i+p_j)^2}
	\left[z_ip_i^\mu-z_jp_j^\mu\right]
	\left[z_ip_i^\nu-z_jp_j^\nu\right]
      \right\}\\
    \langle|\mf{V}_{\gamma s}^a|\rangle
    \,=\;&
      8\pi\mu^{2\epsilon}\alpha
      \left\{
	\frac{2}{2-x_{ij,a}-z_j}-2-\frac{m_s^2}{p_ip_j}
      \right\}
  \end{split}\;.
\end{equation}
Again, only photon splittings exhibit any spin correlations. 
In all other cases the dipole function and the underlying 
Born matrix element factorise.

\subsection{Initial-final dipoles}

Dipoles with the emitter in the initial state and the spectator in the 
final state take the form
\begin{equation}
  \begin{split}
    \mc{D}_{j,k}^a
    \,=\;& -\frac{1}{2p_ap_j}\,\frac{1}{x_{aj,k}}\;\Qop{\ajt\kt}\;
	   {}_m\langle\ldots,\ajt,\ldots,\kt,\ldots|\mf{V}_{j,k}^a|
	              \ldots,\ajt,\ldots,\kt,\ldots\rangle{}_m
  \end{split}
\end{equation}
The charge-correlator is defined in eq.\ \eqref{eq:chargecorrelator}. 
All momenta of the dipole are on-shell 
\begin{equation}
  \begin{split}
    p_a^2=&\;p_\ajt^2=p_i^2=0\,,\qquad p_k^2=p_\kt^2=m_k^2\;,
  \end{split}
\end{equation}
and the total four momentum flowing through it is given by
\begin{equation}
  \begin{split}
    q=&\;-p_a+p_j+p_k=-p_\ajt+p_\kt\;.
  \end{split}
\end{equation}
It is thus invariant under the emission. 
The momenta of the parton before and after the splitting 
are connected through the map 
\begin{equation}
  \begin{split}
    p_\ajt=\;&x_{ai,k}p_a\\
    p_\kt=\;&q+p_\ajt\;.
  \end{split}
\end{equation}
The splitting variables read
\begin{equation}
  \begin{split}
    x_{aj,k}=1-\frac{p_jp_k}{(p_j+p_k)p_a}
    \qquad\text{and}\qquad
    u_j=1-u_k=\frac{p_jp_a}{(p_j+p_k)p_a}\;.
  \end{split}
\end{equation}
As the splitting partons are all in the initial state and 
therefore massless, the singularity at $u_j=0$ is present in any case. 
Adopting the above convention for labeling the spin of the 
emitter $\ajt$ the dipole functions are defined as
\begin{equation}
  \begin{split}
    \langle s|\mf{V}_{\gamma,k}^f|s'\rangle
    \,=\;&
      8\pi\mu^{2\epsilon}\alpha
      \left\{
	\frac{2}{2-x_{aj,k}-u_k}-1-x_{aj,k}-\epsilon(1-x_{aj,k})
      \right\}
      \delta_{ss'}\\
    \langle s|\mf{V}_{f,k}^\gamma|s'\rangle
    \,=\;&
      8\pi\mu^{2\epsilon}\alpha
      \left\{
	1-\epsilon-2x_{aj,k}(1-x_{aj,k})\vphantom{\frac{2}{x_{aj,k}}}
      \right\}
      \delta_{ss'}\\
    \langle\mu|\mf{V}_{f,k}^f|\nu\rangle
    \,=\;&
      8\pi\mu^{2\epsilon}\alpha
      \left\{
	-g^{\mu\nu}x_{aj,k}
	-\frac{1-x_{aj,k}}{x_{aj,k}}\frac{2z_jz_k}{p_jp_k}
	\left[\frac{p_j^\mu}{u_j}-\frac{p_k^\mu}{u_k}\right]
	\left[\frac{p_j^\nu}{u_j}-\frac{p_k^\nu}{u_k}\right]
      \right\}
  \end{split}
\end{equation}
As only massless particles are considered as initial states, 
no splitting functions involving massive scalars are needed.

\subsection{Initial-initial dipoles}

Dipoles with both emitter and spectator in the initinal state take the form
\begin{equation}
  \begin{split}
    \mc{D}_j^{a,b}
    \,=\;& -\frac{1}{2p_ap_j}\,\frac{1}{x_{aj,b}}\;\Qop{\ajt\bt}\;
	   {}_m\langle\ldots,\ajt,\ldots,\bt,\ldots|\mf{V}_j^{a,b}|
	              \ldots,\ajt,\ldots,\bt,\ldots\rangle{}_m
  \end{split}
\end{equation}
The charge-correlator is defined in eq.\ \eqref{eq:chargecorrelator}. 
All momenta of the dipole are on-shell 
\begin{equation}
  \begin{split}
    p_a^2=&\;p_\ajt^2=p_i^2=0\,,\qquad p_b^2=p_\bt^2=0\,,
  \end{split}
\end{equation}
and the total four momentum flowing through it is given by 
\begin{equation}
  \begin{split}
    q=&\;-p_a-p_b+p_j\,,\qquad \qt=-p_\ajt-p_\bt\;.
  \end{split}
\end{equation}
before and after the emission, respectively. 
Although $q$ acquires transverse momentum after the emission, 
$q^2=\qt^2$ holds, i.e.\ the mass of the dipole is invariant. 
The momenta of the partos before and after the splitting 
are connected through the map 
\begin{equation}
  \begin{split}
    p_\ajt=\;&x_{ai,b}p_a\\
    p_\bt=\;&p_b\;.
  \end{split}
\end{equation}
While the initial state cannot absorb recoil transverse to the beam 
access, it is transferred to the final state through a collective 
boost 
\begin{equation}
  \begin{split}
    k_{\tilde{\im}}^\mu
    \,=\;&
      k_i^\nu
      \left[
	\delta_\nu^\mu
	-2\,\frac{(q+\qt)_\nu(q+\qt)^\mu}{(q+\qt)^2}
	+2\,\frac{q_\nu \qt^\mu}{q^2}
      \right]
  \end{split}
\end{equation}
of all final state partons. The splitting variables read 
\begin{equation}
  \begin{split}
    x_{aj,b}=1-\frac{p_j(p_a+p_b)}{p_ap_b}
    \qquad\text{and}\qquad
    v_j=\frac{p_ap_j}{p_ap_b}\;,
    \qquad
    v_b=1\;.
  \end{split}
\end{equation}
As the splitting partons are all in the initial state and 
therefore massless, the singularity at $v_j=0$ is present in any case. 
Adopting the above convention for labeling the spin of the 
emitter $\ajt$ the dipole functions are defined as
\begin{equation}
  \begin{split}
    \langle s|\mf{V}_\gamma^{f,b}|s'\rangle
    \,=\;&
      8\pi\mu^{2\epsilon}\alpha
      \left\{
	\frac{2}{2-x_{aj,b}}-1-x_{aj,b}-\epsilon(1-x_{aj,b})
      \right\}
      \delta_{ss'}\\
    \langle s|\mf{V}_f^{\gamma,b}|s'\rangle
    \,=\;&
      8\pi\mu^{2\epsilon}\alpha
      \left\{
	1-\epsilon-2x_{aj,b}(1-x_{aj,b})\vphantom{\frac{2}{x_{aj,b}}}
      \right\}
      \delta_{ss'}\\
    \langle\mu|\mf{V}_f^{f,b}|\nu\rangle
    \,=\;&
      8\pi\mu^{2\epsilon}\alpha
      \left\{
	-g^{\mu\nu}x_{aj,b}
	-\frac{1-x_{aj,b}}{x_{aj,b}}\frac{2v_jv_b}{p_jp_b}
	\left[\frac{p_j^\mu}{v_j}-\frac{p_b^\mu}{v_b}\right]
	\left[\frac{p_j^\nu}{v_j}-\frac{p_b^\nu}{v_b}\right]
      \right\}
  \end{split}
\end{equation}
As only massless particles are considered as initial states, 
no splitting functions involving massive scalars are needed.

\section{Integrated splitting functions}
\label{app:intsplit}

This appendix summarises the complete functional form of the 
integrated dipole terms cast in the form of the $\Iop$, $\Kop$ 
and $\Pop$ operators \cite{Catani:1996vz,Catani:2002hc}.

\subsection{Terms of the \texorpdfstring{$\Iop$}{I} operator}
Each dipole contribution to the $\Iop$ operator reads
\begin{equation}\label{eq:Iik_app}
  \begin{split}
    \Iop_{ik}(\epsilon,\mu^2;\kappa,\{\alphadip\})
    =\;&
	\Qop{ik}
	\left[
	  \V_{ik}(\epsilon,\mu^2;\kappa)
	  +\Gamma_i(\epsilon,\mu^2)
	  +\gamma_i\left(1+\ln\frac{\mu^2}{s_{ik}}\right)
	  +K_i
	  +A_{ik}^I(\{\alphadip\})
	  +\order(\epsilon)
	\right]\,,\hspace*{-10mm}
  \end{split}
\end{equation}
cf.\ eq.\ \eqref{eq:Iik}. 
Therein, $\V_{ik}$ takes the general form
\begin{equation}
  \begin{split}
    \V_{ik}(\epsilon,\mu^2;\kappa)
    \,=\,\left\{\begin{array}{cl}
	  Q_i^2 
	  \left(\frac{\mu^2}{s_{ik}}\right)^\epsilon
	  \left[\VS_{ik}(\epsilon)+\VNS_{ik}(\kappa)-\tfrac{\pi^2}{3}\right] 
	  & i\neq\gamma\\
	  \mHl \VNS_{\gamma k}(\kappa)
	  & i=\gamma \;.
         \end{array}\right.
  \end{split}
\end{equation}
Wherein, $\VS_{ik}$ contains all infrared singularities (double 
poles for massless fermion splittings, single pole for massive 
splittings), and a non-singular piece $\VNS_{ik}$ that incorporates 
all finite dependences on the masses of the emitter and the spectator. 
The latter vanishes in the massless case, see below.
In the case of photon splittings, only the non-singular part 
survives, again being non-zero only in the case of massive 
spectators.
The singular part vanishes in case of photon splittings. 

The singular term $\VS_{ik}$ is symmetric in $i$ and $k$, 
it is given by
\begin{equation}\label{eq:VS}
  \begin{split}
    \VS_{ik}(\epsilon)
    \,=\;&\frac{1}{v_{ik}}\left(\frac{q_{ik}^2}{s_{ik}}\right)^\epsilon
	  \left[
	    \frac{1}{\epsilon^2}
	    \left(
	      1-\tfrac{1}{2}\rho_i^{-2\epsilon}-\tfrac{1}{2}\rho_k^{-2\epsilon}
	    \right)
	    -\frac{\pi^2}{12}\left(\Theta(m_i)+\Theta(m_k)\right)
	  \right]
  \end{split}
\end{equation}
with
\begin{equation}
  \begin{split}
    \rho_{i/k}
    \,=\;&\sqrt{\frac{(1-v_{ik})s_{ik}+2m_{i/k}^2}{(1+v_{ik})s_{ik}+2m_{i/k}^2}}
	  \stackrel{m_{i/k}\to 0}{\longrightarrow}
	  0
  \end{split}
\end{equation}
and the velocity factors
\begin{equation}
  \begin{split}
    v_{ik}
    \,=\;&\frac{\sqrt{\Kallen{q_{ik}^2}{m_i^2}{m_k^2}}}{s_{ik}}
	  \stackrel{m_i,m_k\to 0}{\longrightarrow}
	  1
  \end{split}
\end{equation}
parametrise the mass dependence. 
Therein, $q_{ik}^2=(p_i+p_k)^2=s_{ik}+m_i^2+m_k^2$ is the dipole 
invariant mass. 
In $d=4-2\epsilon$ dimensions, $\epsilon>0$, 
the massless limit is approached smoothly.
Subsequently taking the limit $\epsilon\to 0$, 
however, results in $\epsilon^{-2}$ poles only if at least one of 
both particles making up the dipole is massless. Completely massive 
dipoles only result in single infrared poles.

The non-singular $\VNS_{ik}$ contributions encode the additional finite 
$m_{i/k}>0$ contributions.
They vanish in the limit that both $m_i\to 0$ and $m_k\to 0$ 
and their precise form can be found in eq.\ (6.21)--(6.26) of 
\cite{Catani:2002hc}. 
Of particular interest is the term for a photon emitter 
in the presence of a massive spectator,
\begin{equation}
  \begin{split}
    \VNS_{\gamma k}(\kappa)
    \,=\;&
      \gamma_\gamma
      \left[
	\log\frac{s_{ik}}{q_{ik}^2}
	-2\log\frac{|q_{ik}|-m_k}{|q_{ik}|}
	-\frac{2m_k}{|q_{ik}|+m_k}
      \right]
      +\frac{\pi^2}{6}-\DiLog{\frac{s_{ik}}{q_{ik}^2}}\\
    &{}
      +2\gamma_\gamma\left(\kappa-\tfrac{2}{3}\right)
       \frac{m_k^2}{s_{ik}}
       \log\frac{2m_k}{|q_{ik}|+m_k}\;,
  \end{split}
\end{equation}
with $|q_{ik}|=\sqrt{q_{ik}^2}$. Thus, choosing $\kappa=\tfrac{2}{3}$ 
somewhat simplifies the integrated subtraction term.

Further,
\begin{equation}\label{eq:Gamma}
  \begin{split}
    \Gamma_i(\epsilon,\mu^2)
    \,=\;
    \left\{\begin{array}{cl}
	     \mHl \frac{1}{\epsilon}\,\gamma_i & i=\text{massless fermion} \\
	     Q_i^2\left[\frac{1}{\epsilon}+\tfrac{1}{2}\ln\frac{m_i^2}{\mu^2}-2\right] & i=\text{massive fermion} \\
	     \mHl \frac{1}{\epsilon}\,\gamma_\gamma & i=\gamma \\ 
	     Q_i^2\left[\frac{1}{\epsilon}+\ln\frac{m_i^2}{\mu^2}-2\right] & i=\text{massive scalar}\;.
           \end{array}\right.
  \end{split}
\end{equation}
In case of the photon, the sum runs over all massless 
charged flavours present in the model. 
The flavour constants $\gamma_i$ and $K_i$ are given by
\begin{equation}\label{eq:flavour-constants-g}
  \begin{split}
    \gamma_i\,=\;
    \left\{\begin{array}{cl}
	      \tfrac{3}{2}Q_i^2 & i=\text{fermion} \\
	      \mHl - \tfrac{2}{3}\sum\limits_f N_{C,f}\,Q_f^2 & i=\gamma \\
	      2Q_i^2 & i=\text{scalar}
           \end{array}\right.
    \qquad\qquad
    K_i\,=\;
    \left\{\begin{array}{cl}
	     \left(\tfrac{7}{2}-\tfrac{\pi^2}{6}\right)Q_i^2 & i=\text{fermion} \\
	     \mHl -\tfrac{10}{9}\sum\limits_f N_{C,f}\,Q_f^2 & i=\gamma \\
	     \left(4-\tfrac{\pi^2}{6}\right)Q_i^2 & i=\text{scalar}
           \end{array}\right.
  \end{split}
\end{equation}

and are independent of the particle masses. 
For the photon, the sums run over all charged massless flavours of the 
model, $N_{C,f}$ being their respective colour multiplicity. 
Lastly, the dependence of the $\Iop$ operator on the technical parameters 
$\{\alphadip\}=\{\alphaII,\alphaIF,\alphaFI,\alphaFF\}$ is encoded in 
the last term $A_{ik}^I(\{\alphadip\})$
which is detailed in App.\ \ref{app:Ialpha}.

\subsection{Terms of the \texorpdfstring{$\Kop$}{K} and \texorpdfstring{$\Pop$}{P} operators}

The $\Kop$ operator of eq.\ \eqref{eq:Kop} reads
\begin{equation}
  \label{eq:Kop_app}
  \begin{split}
    \Kop_{aa'}(x;\{\alphadip\})
    \,=\;&
      \frac{\alpha}{2\pi}
      \left\{
	\Kbar_{aa'}(x)
	-\KFS(x)
	-\sum_i\Qop{ia'}\Kcal_{i,aa'}(x)
	-\sum_k\Qop{a'k}\Kt{aa',k}(x)
      \right.\\
    &\hphantom{\frac{\alpha}{2\pi}\;\;}\left.{}
	-\Qop{a'b}\Ktilde_{aa'}(x)
	+A_{aa'}^K(\{\alphadip\})
	\vphantom{\sum_k}
      \right\}\;.
  \end{split}
\end{equation}
Therein, the $\Kbar$ collect universal terms present 
in all splitting involving an initial state as either 
emitter or spectator. 
They are defined as
\begin{equation}
  \label{eq:Kbar}
  \begin{split}
    \Kbar_{aa'}(x)
    \,=\;&
      \Preg{aa'}(x)\log\frac{1-x}{x}+\Phatp{aa'}(x)\\
    &{}
      +\delta^{aa'}
       \left[
	 Q_a^2\left(\frac{2}{1-x}\,\log\frac{1-x}{x}\right)_+
	 -\delta(1-x)\left(\gamma_a+K_a-\frac{5}{6}\,\pi^2Q_a^2\right)
       \right]\;.
  \end{split}
\end{equation}
As can be seen, they are independent of the final state and 
thus independent of any final state particle masses. 
Here, the regular part of the Altarelli-Parisi splitting functions are 
given by
\begin{equation}
  \label{eq:Pabreg}
  \begin{split}
    \Preg{aa'}(x)
    \,=\;&
    \left\{
      \begin{array}{ll}
	-Q_a^2\, (1+x) & aa'=ff\\
	Q_a^2\, [x^2+(1-x)^2]  & aa'=f\gamma\\
	N_{C,a'}Q_{a'}^2\,\frac{1+(1-x)^2}{x} & aa'=\gamma f\\
	0 & aa'=\gamma\gamma\,,
      \end{array}
    \right.
  \end{split}
\end{equation}
with $N_{C,a'}$ the multiplicity in the QCD representation 
of parton $a'$ (3 for quarks, 1 for leptons).
Similarly, the $\epsilon$-dependent part of the splitting functions 
is defined as 
\begin{equation}
  \label{eq:Pabhat}
  \begin{split}
    \Phatp{aa'}(x)
    \,=\;&
    \left\{
      \begin{array}{ll}
	Q_a^2\, (1-x) & aa'=ff\\
	Q_a^2\, x  & aa'=f\gamma\\
	N_{C,a'}Q_{a'}^2\,x(1-x) & aa'=\gamma f\\
	0 & aa'=\gamma\gamma\,.
      \end{array}
    \right.
  \end{split}
\end{equation}
Thus, for $a=a'=\gamma$ the $\Kbar$ term takes the simple form
\begin{equation}
  \label{eq:Kbargammagamma}
  \begin{split}
    \Kbar_{\gamma\gamma}(x)
    \,=\;& -\tfrac{8}{3}\,\gamma_\gamma\,\delta(1-x)\;,
  \end{split}
\end{equation}
and comprises an end-point term only. 
The factorisation scheme dependent terms $\KFS$ vanish in the \MSbar scheme, 
and the $\Kcal_{i,aa'}$, containing remnants from final state splittings 
where the initial state $a$ forms the spectator. 
As it also depends on the final state flavour $i$, its forms are 
given as follows. 
The contributions, listed separately for scalars, fermions and photons, read
\begin{equation}
  \label{eq:Kcal_f}
  \begin{split}
    \Kcal_{f,\gamma f}(x)
    \,=\;&
      0\\
    \Kcal_{f,ff}(x)
    \,=\;&
      2
      \left[
	\left(\frac{\log(1-x)}{1-x}\right)_+
	-\frac{\log(2-x)}{1-x}
      \right]
      +\left[J_{\gamma Q}^a\left(x,\frac{m_i}{\sqrt{s_{ik}}}\right)\right]_+\\
      &{}
      +2\left(\frac{1}{1-x}\right)_+\log\frac{(2-x)s_{ik}}{(2-x)s_{ik}+m_i^2}
      +\delta(1-x)
      \left[
	-\frac{\gamma_f}{Q_i^2}
	+\frac{m_i^2}{s_{ik}}\,\log\frac{m_i^2}{s_{ik}+m_i^2}
	+\tfrac{1}{2}\frac{m_i^2}{s_{ik}+m_i^2}
      \right]
      \\
    \Kcal_{f,f\gamma}(x)
    \,=\;&
      2\,\frac{m_i^2}{xs_{ik}}\,\log\frac{m_i^2}{(1-x)s_{ik}+m_i^2}\\
    \Kcal_{f,\gamma\gamma}(x)
    \,=\;& \Kcal_{f,ff}\;,
  \end{split}
\end{equation}
and
\begin{equation}
  \label{eq:Kcal_s}
  \begin{split}
    \Kcal_{s,aa'}(x)
    \,=\;&
      \Kcal_{f,aa'}(x)
      -\delta^{aa'}
       \left[
         \left(\frac{s_{ik}^2(1-x)}{2[s_{ik}(1-x)+m_i^2]^2}\right)_+
         -\delta(1-x)
         \left(
           \frac{m_i^2}{s_{ik}}\log\frac{m_i^2}{q_{ik}^2}+\frac{m_i^2}{2q_{ik}^2}
           +\frac{\gamma_s-\gamma_f}{Q_s^2}
         \right)
       \right]\;.
  \end{split}
\end{equation}
The missing function $J_{\gamma Q}^a(x,y)$ is given in eq.\ (5.58) of 
\cite{Catani:2002hc}.
In the massless limit this reduces to
\begin{equation}
  \begin{split}
    \Kcal_{i,aa'}(x)
    \,=\;&
      -\delta^{aa'} \frac{\gamma_i}{Q_i^2}
       \left[\left(\frac{1}{1-x}\right)_++\delta(1-x)\right]
  \end{split}
\end{equation}
Simultaneously, for $i=\gamma$
\begin{equation}
  \label{eq:Kcal_gamma}
  \begin{split}
    \Kcal_{\gamma,aa'}(x)
    \,=\;&
      -\delta^{aa'}\gamma_\gamma\left[\left(\frac{1}{1-x}\right)_++\delta(1-x)\right]
  \end{split}
\end{equation}
due to the different normalisation of $\Qop{ik}$ in the photon case.
The $\Kt{aa',k}$ terms comprise of initial state splittings in the presence 
of a final state spectator. 
They take the form
\begin{equation}
  \label{eq:Kt}
  \begin{split}
    \Kt{aa',k}(x)
    \,=\;&
      \Preg{aa'}(x)\,\log\frac{(1-x)s_{ak}}{(1-x)s_{ak}+m_k^2}\\
    &{}
      +\gamma_a\,\delta^{aa'}\,\delta(1-x)
       \left[
	 \log\frac{s_{ak}-2m_k\sqrt{s_{ak}+m_k^2}+2m_k^2}{s_{ak}}
	 +\frac{2m_k}{\sqrt{s_{ak}+m_k^2}+m_k}
       \right]
  \end{split}
\end{equation}
Again, for $a=a'=\gamma$ the end-point contribution cannot arise as 
no corresponding splitting is integrated over, $\Kt{\gamma\gamma,k}=0$.
Finally, the $\Ktilde$ terms, containing correlations between both 
initial state partons, read
\begin{equation}
  \label{eq:Ktilde}
  \begin{split}
    \Ktilde_{aa'}(x)
    \,=\;&
      \Preg{aa'}(x)\,\log(1-x)
      +Q_{a'}^2\delta^{aa'}\left[2\left(\frac{\log(1-x)}{1-x}\right)_+-\frac{\pi^2}{3}\,\delta(1-x)\right]\;.
  \end{split}
\end{equation}
They thus vanish for $a=a'=\gamma$.

Finally, the $\Pop$ operator is given by 
\begin{equation}
  \label{eq:Pop_app}
  \begin{split}
    \Pop_{aa'}(x,\mu_F^2)
    \,=\;&
      \frac{\alpha}{2\pi}\;P^{aa'}(x)
      \left[
	\sum\limits_k\Qop{a'k}\,\log\frac{\muF^2}{xs_{ak}}
	+\Qop{a'b}\,\log\frac{\muF^2}{xs_{ab}}
      \right]\;,
  \end{split}
\end{equation}
cf.\ eq.\ \eqref{eq:Pop}.
Therein, the regularised Altarelli-Parisi splitting functions are given by
\begin{equation}
  \label{eq:Pab}
  \begin{split}
    P^{aa'}(x)
    \,=\;&
      \Preg{aa'}(x)
      +\delta^{aa'}
       \left[
	 2\,Q_a^2\left(\frac{1}{1-x}\right)_++\gamma_a\,\delta(1-x)
       \right]\;.
  \end{split}
\end{equation}
Most importantly, 
\begin{equation}
  \begin{split}
    P^{\gamma\gamma}(x)
    \,=\;&
      \gamma_\gamma\,\delta(1-x)\;.
  \end{split}
\end{equation}

\section{Precise forms of \texorpdfstring{$A_{ik}^I$}{AikI} and \texorpdfstring{$A_{ab}^K$}{AabK}}
\label{app:Ialpha}

\subsection{The \texorpdfstring{$A_{ik}^I$}{AikI} term}

The explicit dependence of the $\Iop$ operator on the $\{\alphadip\}$ 
parameters, or more precisely only on $\alphaFF$, is given by 
\cite{Nagy:1998bb,Nagy:2003tz,Frederix:2010cj,Campbell:2004ch,Bevilacqua:2009zn}
if $i$ is a massless fermion
\begin{equation}\label{eq:AalphaFF-f}
  \begin{split}
    A_{ik}^I(\{\alpha\})
    \,=\;&
	  -\ln^2\alphaFF-\gamma_i\left(\ln\alphaFF+1-\alphaFF\right)
	  \qquad\qquad m_k=0 \\
    \,=\;&
	  \ln^2\frac{1-y_+^2+2xy_+}{(1-x+y_+)(1+x-y_+)}
	  -2\ln^2\frac{1+y_+-x}{1+y_+}\\
	 &{}
	  +4\left[
	      \ln\frac{1+y_+}{2}\ln\frac{1-y_++x}{1-y_+}
	      +\ln\frac{1+y_+}{2y_+}\ln\frac{1-y_+^2+2x_+y_+}{1-y_+^2}
	    \right.\\
	 &\left.\hspace*{10mm}{}
	      +\DiLog{\frac{1-y_+}{1+y_+}}
	      -\DiLog{\frac{1-y_+}{2}}
	      +\DiLog{\frac{1+x-y_+}{2}}
	      -\DiLog{\frac{1-y_+^2+2xy_+}{(1+y_+)^2}}
	    \right]\\
	 &{}
	  -\tfrac{3}{2}\left(\ln\alphaFF+y_+(1-\alphaFF)\right)
	  \qquad\qquad m_k\neq 0 \\
  \end{split}
\end{equation}
or if $i$ a massive fermion
\begin{equation}\label{eq:AalphaFF-fm}
  \begin{split}
    A_{ik}^I(\{\alpha\})
    \,=\;&
	  2\left[
	     -\ln\alphaFF\left(1+\ln\mu_i^2\right)
	     +\DiLog{\alphaFF\frac{\mu_i^2-1}{\mu_i^2}}
	     -\DiLog{\frac{\mu_i^2-1}{\mu_i^2}}
	   \right]\\
	 &{}
	  +\tfrac{1}{2}\left[
	     -(1-\alphaFF)\frac{3\alphaFF(1-\mu_i^2)+2\mu_i^2}
	                       {\alphaFF+(1-\alphaFF)\mu_i^2}
	     +\frac{1+\mu_i^2}{1-\mu_i^2}
	      \ln\left(\alphaFF+(1-\alphaFF)\mu_i^2\right)
	   \right]
	   \\
	 &{}
	  \qquad m_k=0 \\
    \,=\;&
          \tfrac{3}{2}(1+\alphaFF y_+)
          +\frac{1}{1-\mu_k}
          -\frac{2-2\mu_i^2-\mu_k}{d}
          +\tfrac{1}{2}\frac{(1-\alphaFF y_+)\mu_i^2}{\mu_i^2+2\alphaFF y_+d}
          -2\ln\alphaFF
	   \\
	 &{}
          +\tfrac{1}{2}\frac{\mu_i^2+d}{d}
           \ln\frac{\mu_i^2+2\alphaFF y_+d}{(1-\mu_k)}
	   \\
	 &{}
	  +\frac{2}{v_{ik}}\left[{}
	      -\DiLog{\frac{a+x}{a+x_+}}
	      +\DiLog{\frac{a}{a+x_+}}
	      +\DiLog{\frac{x_--x}{x_-+a}}
	      -\DiLog{\frac{x_-}{x_-+a}}
	   \right.\\
	 &\left.\hspace*{11mm}{}
	      +\DiLog{\frac{c+x}{c+x_+}}
	      -\DiLog{\frac{c}{c+x_+}}
	      -\DiLog{\frac{x_--x}{x_-+c}}
	      +\DiLog{\frac{x_-}{x_-+c}}
	   \right.\\
	 &\left.\hspace*{11mm}{}
	      -\DiLog{\frac{b-x}{b-x_-}}
	      +\DiLog{\frac{b}{b-x_-}}
	      +\DiLog{\frac{x_+-x}{x_+-b}}
	      -\DiLog{\frac{x_+}{x_+-b}}
	   \right.\\
	 &\left.\hspace*{11mm}{}
	      +\DiLog{\frac{b-x}{b+a}}
	      -\DiLog{\frac{b}{b+a}}
              -\DiLog{\frac{c+x}{c-a}}
              +\DiLog{\frac{c}{c-a}}
	   \right.\\
	 &\left.\hspace*{11mm}{}
              +\ln(c+x) \ln\frac{(a-c)(x_+-x)}{(a+x)(x_++c)}
              -\ln c \ln\frac{(a-c)x_+}{a(x_++c)}
	   \right.\\
	 &\left.\hspace*{11mm}{}
              -\ln(b-x) \ln\frac{(a+b)(x_--x)}{(a+x)(x_--b)}
              +\ln b \ln\frac{(a+b)x_-}{a(x_--b)}
	   \right.\\
	 &\left.\hspace*{11mm}{}
              -\ln\left((a+x)(b-x_+)\right) \ln(x_+-x)
              +\ln\left(a(b-x_+)\right) \ln x_+
	   \right.\\
	 &\left.\hspace*{11mm}{}
              +\ln d \ln\frac{(a+x)x_+x_-}{a(x_+-x)(x_--x)}
              +\ln\frac{x_--x}{x_-} \ln\frac{c+x_-}{a+x_-}
	   \right.\\
	 &\left.\hspace*{11mm}{}
              +\tfrac{1}{2}\ln\frac{a+x}{a}\ln\left(a(a+x)(a+x_+)^2\right)
            \right]
	  \qquad m_k\neq 0 \\
  \end{split}
\end{equation}
or if $i$ is a photon
\begin{equation}\label{eq:AalphaFF-a}
  \begin{split}
    \,=\;&
	  -\gamma_i\left(\ln\alphaFF+1-\alphaFF\right)
	  \qquad\qquad m_k=0 \\
    \,=\;&
	  -\gamma_i
	  \left[\left(\frac{1-\mu_k-\alphaFF y_+(1+\mu_k)}{1+\mu_k}
				  +\ln\frac{\alphaFF y_+(1+\mu_k)}{1-\mu_k}\right)
	   \right.\\
	 &\left.\hspace*{16mm}{}
                +\tfrac{3}{2}(\kappa-\tfrac{2}{3})\frac{2\mu_k^2}{1-\mu_k^2}
                 \ln\frac{(1-\alphaFF y_+)(1+\mu_k)}{2\mu_k}
          \right]
	  \qquad\qquad m_k\neq 0 \\
  \end{split}
\end{equation}
or a massive scalar
\begin{equation}\label{eq:AalphaFF-s}
  \begin{split}
    \,=\;&
	  2\left[
	    -\ln\alphaFF(1+\ln\mu_i^2)
	    +\DiLog{\alphaFF\frac{\mu_i^2-1}{\mu_i^2}}
	    -\DiLog{\frac{\mu_i^2-1}{\mu_i^2}}
	    -(1-\alphaFF)
	      \frac{\alphaFF(1-\mu_i^2)+\mu_i^2}
	           {\alphaFF+(1-\alphaFF)\mu_i^2}
	   \right]
	  \qquad m_k=0 \hspace*{-20mm}\\
    \,=\;&
	  \tfrac{3}{2}(1+\alphaFF y_+)
	  +\frac{1}{1-\mu_k}
          -\frac{2-2\mu_i^2-\mu_k}{d}
          +\tfrac{1}{2}\frac{(1-\alphaFF y_+)\mu_i^2}{\mu_i^2+\alphaFF y_+ d}
          -2\ln\alphaFF
          \\
         &{}
	  -\tfrac{1}{2}(1-2\alphaFF)y_+
	   \frac{\alphaFF y_+ d (1-\mu_k)+\mu_i^2}
                {(2\alphaFF y_+ d+\mu_i^2)(1-\mu_k)}
          \\
	 &{}
	  +\frac{2}{v_{ik}}\left[{}
	      -\DiLog{\frac{a+x}{a+x_+}}
	      +\DiLog{\frac{a}{a+x_+}}
	      +\DiLog{\frac{x_--x}{x_-+a}}
	      -\DiLog{\frac{x_-}{x_-+a}}
	   \right.\\
	 &\left.\hspace*{11mm}{}
	      +\DiLog{\frac{c+x}{c+x_+}}
	      -\DiLog{\frac{c}{c+x_+}}
	      -\DiLog{\frac{x_--x}{x_-+c}}
	      +\DiLog{\frac{x_-}{x_-+c}}
	   \right.\\
	 &\left.\hspace*{11mm}{}
	      -\DiLog{\frac{b-x}{b-x_-}}
	      +\DiLog{\frac{b}{b-x_-}}
	      +\DiLog{\frac{x_+-x}{x_+-b}}
	      -\DiLog{\frac{x_+}{x_+-b}}
	   \right.\\
	 &\left.\hspace*{11mm}{}
	      +\DiLog{\frac{b-x}{b+a}}
	      -\DiLog{\frac{b}{b+a}}
              -\DiLog{\frac{c+x}{c-a}}
              +\DiLog{\frac{c}{c-a}}
	   \right.\\
	 &\left.\hspace*{11mm}{}
              +\ln(c+x) \ln\frac{(a-c)(x_+-x)}{(a+x)(x_++c)}
              -\ln c \ln\frac{(a-c)x_+}{a(x_++c)}
	   \right.\\
	 &\left.\hspace*{11mm}{}
              -\ln(b-x) \ln\frac{(a+b)(x_--x)}{(a+x)(x_--b)}
              +\ln b \ln\frac{(a+b)x_-}{a(x_--b)}
	   \right.\\
	 &\left.\hspace*{11mm}{}
              -\ln\left((a+x)(b-x_+)\right) \ln(x_+-x)
              +\ln\left(a(b-x_+)\right) \ln x_+
	   \right.\\
	 &\left.\hspace*{11mm}{}
              +\ln d \ln\frac{(a+x)x_+x_-}{a(x_+-x)(x_--x)}
              +\ln\frac{x_--x}{x_-} \ln\frac{c+x_-}{a+x_-}
	   \right.\\
	 &\left.\hspace*{11mm}{}
              +\tfrac{1}{2}\ln\frac{a+x}{a}\ln\left(a(a+x)(a+x_+)^2\right)
            \right]
	  \qquad\qquad m_k\neq 0
  \end{split}
\end{equation}
As can be seen, the massless case can be recovered from the massive 
one in the limit that $m_k\to 0$.
The $A_{ik}^I$ term could in principle also be shifted into the $\Kop$ 
operator to leave the $\Iop$ operator invariant. 
To be applicable to pure final-state calculations, though, it is left 
where it is.
The various abbreviations above are defined as
\begin{equation}
  \begin{split}
    \mu_{i/k}^2\,=\;&\frac{m_{i/k}^2}{q_{ik}^2}\\
    y_+\,=\;&\frac{(1-\mu_k)^2-\mu_i^2}{1-\mu_i^2-\mu_k^2} 
	    \,=\;
	    \left\{
	      \begin{array}{cl}
		1 & \text{if } m_k=0 \\
		\frac{1-\mu_k}{1+\mu_k} & \text{if } m_i=0
	      \end{array}
	    \right.\\
    y\,=\;&\tfrac{1}{2}\left[1-\mu_k^2+\alphaFF(1-\mu_k)^2
                             -(1-\mu_k)\sqrt{(\alphaFF(1-\mu_k))^2+(1+\mu_k)^2
                                             -2\alphaFF(1+\mu_k^2)}\right]\\
    x\,=\;&(1-\alphaFF)y_+
           +\sqrt{(1-\alphaFF)
		  \left(1 - \alphaFF y_+^2
			-\frac{4\mu_i^2\mu_k^2}{d^2}\right)}\\
    x_\pm\,=\;&\tfrac{1}{4}y_+\pm\tfrac{1}{2}\frac{\sqrt{\Kallen{1}{\mu_i^2}{\mu_k^2}}}{d}\\
    a\,=\;&\frac{\mu_k}{d}\\
    b\,=\;&\frac{1-\mu_k}{d}\\
    c\,=\;&abd\,=\;1-y_+\\
    d\,=\;&\tfrac{1}{2}\frac{s_{ik}}{q_{ik}^2}\,=\;\tfrac{1}{2}(1-\mu_i^2-\mu_k^2)\\
  \end{split}
\end{equation}

\subsection{The \texorpdfstring{$A_{ab}^K$}{AabK} term}

The dependence of the $\Kop$ operator on the $\{\alphadip\}$ 
parameters, or more precisely only on $\alphaFI$, $\alphaIF$ and 
$\alphaII$, is given in  \cite{Nagy:1998bb,Nagy:2003tz,Frederix:2010cj,Campbell:2004ch,Campbell:2005bb}.
It has the general form
\begin{equation}
  \begin{split}
    A_{ab}^K(x,\{\alphadip\})
    \,=\;&
      A_{ab}^{\Kt{}}(x,\alphaFI,\alphaIF)+A_{ab}^{\Ktilde}(x,\alphaII)\;.
  \end{split}
\end{equation}
The superscripts indicate the general functional form of the 
contribution.
In case of the $\alphaFI$ and $\alphaIF$ depedence the existing 
plus distributions are modified in the $\Kt{}$ terms as follows 
\begin{equation}
  \begin{split}
    \left(..\right)_+ \to \left(..\right)_{1-\alphaFI}
  \end{split}
\end{equation}
with
\begin{equation}
  \begin{split}
    \int\limits_0^1\done x\; f(x) [g(x)]_{1-\alphaFI}
    \;=\;&
      \int\limits_{1-\alphaFI}^1\done x\; g(x) [f(x)-f(1)]\;.
  \end{split}
\end{equation}
As the singular point sits at $x=1$ this modification amouts to 
a finite contribution.
The $\alphaIF$ on the other hand leads to finite regular functional 
contribution in the $A_{ab}^{\Kt{}}$ term given in 
\cite{Campbell:2004ch,Campbell:2005bb}.

Lastly, the $\alphaII$ dependence reads \cite{Nagy:2003tz}
\begin{equation}
  \begin{split}
    A_{ab}^{\Ktilde}(x,\alphaII)
    \,=\;
      \log\frac{\alphaII}{1-x}\;,
  \end{split}
\end{equation}
except when $a=b=\gamma$, in which case it is zero as the original 
$\Ktilde_{\gamma\gamma}$ term.

%% file: master.bbl
\begin{mcitethebibliography}{10}
\bibitem{KallweitMunich}
{\sc Munich} is the abbreviation of ``MUlti-chaNnel Integrator at Swiss (CH)
  precision'' -- an parton level NLO generator by S.~Kallweit. In preparation.%
\relax\mciteBstWouldAddEndPunctfalse
\mciteSetBstMidEndSepPunct{\mcitedefaultmidpunct}
{}{\mcitedefaultseppunct}\relax
\EndOfBibitem
\end{mcitethebibliography}
